\documentclass[11pt,a4paper,oneside]{scrarticle}
\usepackage{filecontents}
\pdfoutput=1

\usepackage[tmargin=2.5cm, bmargin=2.5cm, lmargin=2.5cm, rmargin=2.5cm]{geometry}

\usepackage[utf8]{inputenc}
\usepackage[english]{babel}

\usepackage[font={small},labelfont=bf]{caption} 
\setcapindent{0pt} 

\usepackage{csquotes}
\usepackage{courier}

\usepackage{textcomp} 
\usepackage{mathtools}
\usepackage{subfig}

\usepackage{amsmath}
\usepackage{amssymb}

\usepackage[export]{adjustbox} 
\usepackage{here}

\usepackage[singlespacing]{setspace}

\usepackage{tabularx} 

\usepackage{graphicx}
\usepackage{xr} 
\usepackage{nameref}
\usepackage[style=phys,biblabel=brackets]{biblatex} 
\bibliography{main.bib}
\renewbibmacro*{note+pages}{%
	\printfield{note}%
	\setunit{\bibpagespunct}%
	\printfield{pages}%
}

\usepackage{url} 
\usepackage{hyperref} 
\hypersetup{colorlinks, anchorcolor = black, citecolor = blue, urlcolor=blue,
	filecolor=green, linkcolor=black, plainpages=false}

\usepackage[automark,headsepline]{scrlayer-scrpage} 
\clearpairofpagestyles
\lefoot[\pagemark]{\pagemark}
\rofoot[\pagemark]{\pagemark}

\lehead{\headmark}
\rohead{\headmark}
\usepackage{authblk}
\title{A combined diffusion/rate equation model to describe charge generation in phase-separated donor-acceptor blends}
\author[1]{Phillip Teschner}
\author[1]{Atul Shukla}
\author[1]{Dieter Neher}
\affil[1]{Institute of Physics and Astronomy, University of Potsdam, Karl-Liebknecht Straße 24/25, 14476 Potsdam, Germany}
\date{}
\newcommand{\etal}{$et~al.$\ }

\begin{document}
\maketitle

\section*{Abstract}
Over the past decade, the power conversion efficiency (PCE) of organic solar cells (OSCs) has been largely improved by the introduction of novel non-fullerene acceptors (NFAs). Further improvements in PCE require a more comprehensive understanding of the free charge generation process. Recently, the small PCE of donor-acceptor blends with low offsets between the relevant frontier orbitals was attributed to inefficient exciton dissociation. However, another source of photocurrent loss is the competition between exciton diffusion and decay, which is particularly relevant for bilayers or bulk heterojunction blends with large and pure acceptor and/or donor domains. Here, we present an analytical model that combines exciton diffusion with a set of rate equations based on Marcus theory of charge transfer. The model is capable of describing the steady-state and transient charge generation characteristics, while simultaneously incorporating the size and shape of the acceptor/donor domains. An expression for the charge generation efficiency is derived from the steady-state solution of the model. Thereby, the intrinsic exciton lifetime is identified as a pivotal parameter to facilitate efficient charge generation in spite of a vanishing driving force for exciton dissociation. The dynamic formulation of the model is employed to elucidate the characteristic time scales of charge generation. It is found that for low-offset systems, the pure diffusive times are considerably shorter than those associated with charge generation. It can therefore be concluded that when estimating domain sizes via exciton diffusion measurements, the assumption that excitons are instantaneously quenched at the donor-acceptor interface is only valid when a high driving force for exciton dissociation is present. Finally, the model is applied to the results of transient absorption measurements on a PM6:Y6 blend. It is demonstrated that the charge generation dynamics are determined by the interplay between exciton diffusion and hole transfer kinetics, with an estimated Y6 domain size of $\approx 25\,$nm, while interfacial charge transfer (CT) states separate rapidly into free charges.

\section{Introduction}
Recent years have witnessed a consistent improvement in the device performance of bulk heterojunction (BHJ) based organic solar cells (OSCs), with power conversion efficiencies (PCEs) now reaching $20\,$\% \cite{Liu.2024}\cite{Jiang.2024}\cite{Chen.2024}\cite{Guan.2024}. Further improvements in PCE require a critical understanding of the charge generation process and the elimination of all loss pathways. The impressive rise in performance of OSCs has been driven by the development of narrow-gap non-fullerene acceptors (NFAs) \cite{Armin.2021}. Currently, state-of-the-art NFA-based BHJs exhibit high photon absorption across the solar spectrum, coupled with a near unity absorbed-photon-to-charge conversion efficiency, ultimately resulting in a high short-circuit current ($J_\mathrm{SC}$) exceeding $25\,\frac{\mathrm{mA}}{\mathrm{cm}^2}$ \cite{Jia.2023}\cite{Shoaee.2024}. However, OSCs still fall behind their inorganic and perovskite counterparts in terms of the open-circuit voltage ($V_\mathrm{OC}$) \cite{Guan.2024}\cite{Brinkmann.2022}\cite{Zhang.2023}. A crucial step for free charge generation in OSCs is the charge transfer from photo-excited donor or acceptor materials across the interface, which is driven by the energy offset ($\Delta E_\mathrm{LE-CT}$) between the local singlet exciton (LE) in donor/acceptor materials and the charge transfer (CT) state. However, this energy offset, often referred to as the driving force, contributes significantly to the voltage loss in OSCs \cite{Ward.2015}. Therefore, it is crucial to understand and promote routes to achieve efficient free charge generation with minimal driving force in order to maximise both photocurrent and voltage. \\
Free charge generation in OSCs is a multi-step process. The initial step involves diffusion of the photo-generated excitons to the donor-acceptor interface and their subsequent dissociation into a CT state, also referred to as the charge generation step. This is followed by splitting of the CT state into unbound charge carriers, commonly referred to as charge separated (CS) state. In NFA-based BHJs, hole transfer (HT) from the LE state of the NFA has been identified as the dominant charge generation pathway. In the framework of semi-classical Marcus charge transfer model, reducing the driving force can critically limit the HT rate, which ultimately hinders the overall photocurrent generation efficiency in low-offset NFA-based OSCs \cite{Karuthedath.2021}\cite{Nakano.2019}\cite{Vandewal.2012}\cite{Vandewal.2010}. In $2021$, Zhou \etal demonstrated a reduction of more than two orders of magnitude in the HT rate, as observed through transient absorption spectroscopy (TAS), when $\Delta E_\mathrm{LE-CT}$ was decreased from $0.5\,$eV to $0\,$eV, consistent with the predictions of Marcus theory \cite{Zhou.2021}. Herein, the authors used dilute acceptor blends with donor:acceptor ratio of $4$:$1$ to directly probe the intrinsic HT dynamics with negligible contribution from bulk exciton diffusion. However, in optimised BHJs, morphology - encompassing domain sizes, crystalline order, and phase separation - plays a critical role in device performance alongside energetics, as it influences exciton utilisation, charge separation and carrier transport properties. Another important observation is that diminishing values of $\Delta E_\mathrm{LE-CT}$ can facilitate a reverse transition from the CT to the LE state, leading to reformation of singlet excitons \cite{Classen.2020}\cite{Riley.2022}\cite{Sun.2023}. More recently, Pranav \etal used a steady-state rate model in conjunction with Marcus theory and density functional theory simulations, considering the combined role of exciton dissociation and reformation to successfully describe the free charge generation and photoluminescence quantum yield (PLQY) in several OSCs with varying energy-level offset \cite{Pranav.2024}. In this work, the authors also demonstrated a rather sharp decline in free charge generation efficiency with minor changes in $\Delta E_\mathrm{LE-CT}$ values, setting a critical limit on the permissible offset value to simultaneously maximise photovoltage and photocurrent in modern NFA-based OSCs. \\ 
On the theoretical side, the complete prediction of the charge generation process in BHJs can be complex. The current deterministic models do not completely encompass the role of parameters such as exciton diffusion, exciton lifetime, domain size and interfacial energetics, as well as their mutual interactions. For example, de Falco \etal developed a multiscale drift-diffusion model for OSCs that encompasses a free charge generation model including exciton diffusion. However, they did not investigate the impact of interfacial energetics on free charge generation \cite{Falco.2012}. The rate equation model employed by Classen \etal and later Pranav \etal implicitly accounts for exciton diffusion solely by means of a reduction factor (set to 0.1) in the exciton dissociation rate \cite{Classen.2020}\cite{Pranav.2024}. In another work, Riley \etal considered the role of one-dimensional exciton diffusion in their charge generation model, but did not explicitly consider the charge generation rates of Marcus theory \cite{Riley.2022}. While these models can reproduce charge generation efficiencies in several BHJs, these analyses may render inaccuracies in certain systems. For example, parameters such as poor exciton diffusion or large domain sizes may impose an intrinsic limit to the charge generation efficiencies \cite{Riley.2023}. Further, a deeper microscopic understanding also requires time-dependent modelling of charge generation that can be applied to transient and quasi-steady state techniques such as TAS \cite{Price.2022}, time-resolved photoluminescence \cite{Gasparini.2021} and pulsed-PLQY \cite{Riley.2023}\cite{Riley.2022b}. \\
In this work, we develop a differential equation model that combines an exciton diffusion model with a rate equation model at the donor-acceptor interface. The model is capable of describing both charge generation under constant illumination and the dynamics of charge generation, thereby enabling the simulation of time-resolved experiments. We begin with the formulation of our differential equation model for an arbitrary three-dimensional geometry of donor/acceptor domains. We then proceed by considering spherical and planar geometries to address specific heterojunction designs in OSCs. Subsequently, the steady-state formulation of the model is employed to derive an analytic expression of the charge generation efficiency, which is then evaluated as a function of $\Delta E_\mathrm{LE-CT}$ using Marcus theory. We compare our findings with the results obtained from an established rate model, in which diffusion is included implicitly by means of a reduction factor in the exciton dissociation rate. Our efficiency exhibits minor yet significant differences compared to the outcomes from the steady-state rate model reported earlier. Furthermore, the impact of various parameters is discussed, with a special emphasise on the relevance of the intrinsic exciton lifetime. We continue by simulating a TAS-like experiment, in which the acceptor is selectively excited, in order to unravel the characteristic times of charge generation. This allows us to elucidate the impact of singlet exciton diffusion, dissociation and reformation on the temporal evolution of the ground state bleach (GSB) of the donor and of electro-absorption (EA), the latter being generally used to track charge separation. In particular, we will show that for low-offset systems, the timescale of pure diffusion to the donor-acceptor interface can be significantly shorter than the timescale of charge generation. Finally, we will compare our model results with experimental data on PM6:Y6. The observation of long timescales ($\sim 100\,$ps) required for the completion of HT to the donor agrees with the timescales predicted by our model. Moreover, we utilize the model to estimate the average domain size of Y6, which is found to be within the range of reported literature values. 

\section{Modelling} 
\label{sec:modelling}
We consider a pure acceptor domain, which we describe by some compact set $\Omega \subset \mathbb{R}^3$. The following considerations can be equally applied to a donor domain in contact with an acceptor. The boundary $\Gamma \coloneq \partial \Omega$ is assumed to be piecewise smooth and orientable, so that we can apply the divergence theorem on $\Omega$. Moreover, the boundary is decomposed into two types, $\Gamma = \Gamma_\mathrm{d}\,\dot\cup\,\Gamma_\mathrm{r}$. $\Gamma_\mathrm{d}$ models a boundary that enables exciton dissociation (e.g.\ the interface with a suitable donor domain) and $\Gamma_\mathrm{r}$ implements a reflecting boundary (e.g.\ the interface with glass). Further, $\boldsymbol{\nu}: \Gamma \rightarrow \mathbb{R}^3$ denotes the outward unit normal vector on $\Gamma$.
\begin{figure}[t!]
	\includegraphics[width = 1 \textwidth]{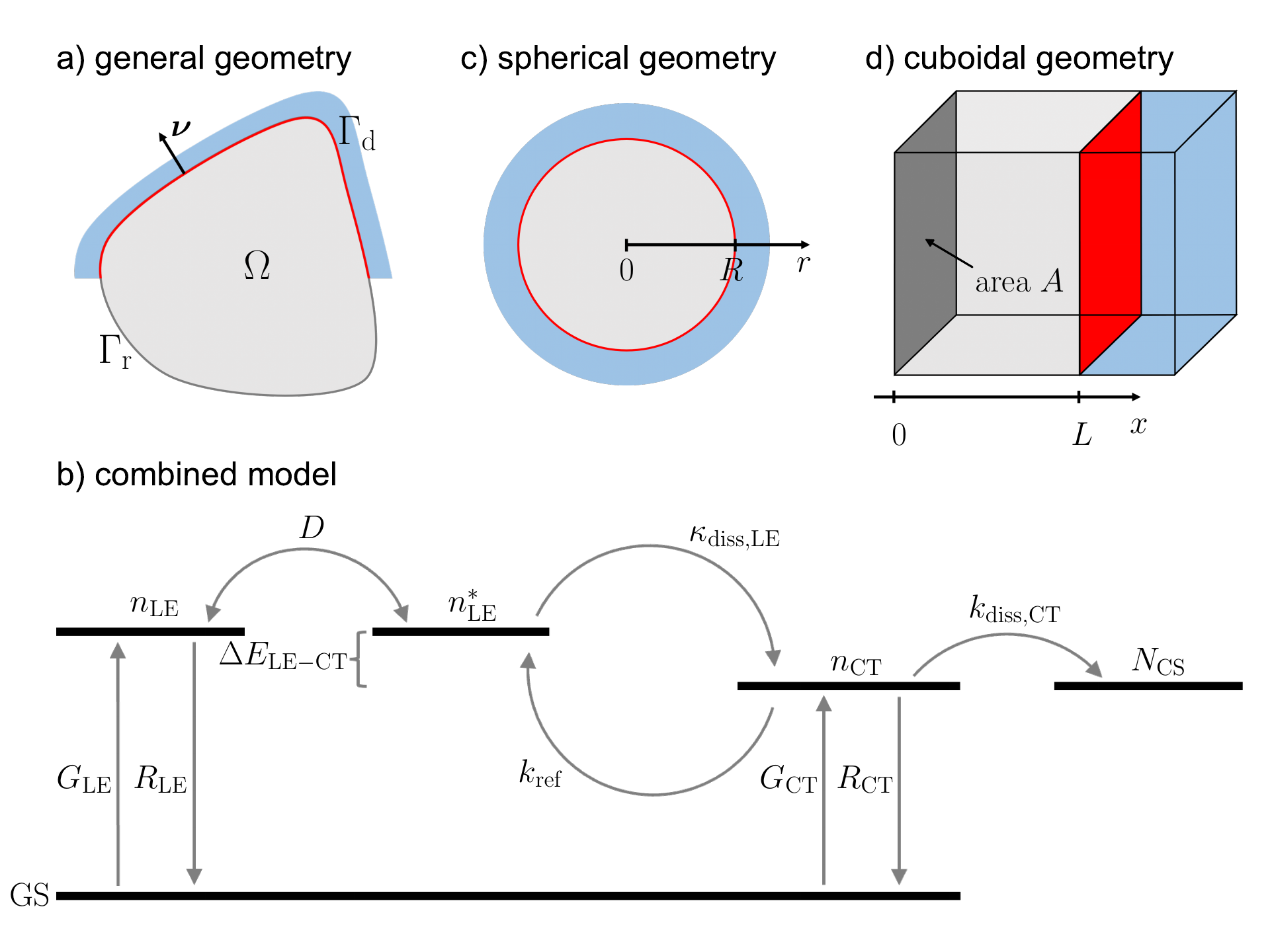}
	\caption{\textbf{a)} General geometry of the model. $\Omega$ denotes the set attributed to the acceptor domain. $\Gamma_\mathrm{d}$ (red) describes the subset of $\partial \Omega$ that enables exciton dissociation while $\Gamma_\mathrm{r}$ (dark grey) models a reflecting boundary. $\boldsymbol{\nu}$ denotes the outward unit normal vector on $\partial \Omega$. The complementary donor domain, although not explicitly included in our model, is indicated by the blue shaded area. \textbf{b)} Schematic illustration of the combined charge generation model, consisting of a diffusion and a rate equation model. Exciton diffusion in the bulk is modelled by the exciton diffusion equation, which is indicated by the diffusivity $D$. The rate equation model is implemented at the boundary $\Gamma_\mathrm{d}$. $n_\mathrm{LE}^{\ast}$ represents the exciton density at the interface from which the CT state can be formed ($n_\mathrm{CT}$). The CT state can then either reform a singlet exciton, recombine or dissociate into free charges ($N_\mathrm{CS}$). The respective generation ($G$) and recombination rates ($R$) are also shown. \textbf{c)} Spherical geometry to model a BHJ. The sample acceptor domain is completely surrounded by a matching donor, which renders the domain boundary favourable for exciton dissociation. \textbf{d)} Cuboidal geometry to account for a bilayer OSC. The plain $x=L$ is a donor-acceptor interface while the other boundaries are assumed to be reflective.}
	\label{fig:modelling}
	\centering
\end{figure}
The general geometry of the model is summarised in figure \textbf{\ref{fig:modelling}a)}. A donor domain, although not explicitly included in our model, is indicated by the blue coloured region adjacent to $\Gamma_\mathrm{d}$.\\
We describe the excitons in our sample domain $\Omega$ in a deterministic continuum picture with a volume density $n_\mathrm{LE}(\mathbf{x},t)$. The sample domain is regarded as an average over the whole system which incorporates a substantial number of excitons under illumination (similarly, experiments such as TAS also average over different local configurations). Thus, the deterministic continuum approach is valid \cite{Erban.2007}.
Exciton motion in organic solids is understood to be a hopping process facilitated by Förster or Dexter energy transfer \cite{Mikhnenko.2015}\cite{Menke.2014}. For temperatures above $150\,\mathrm{K}$, this hopping can be considered a diffusion process \Cite{Mikhnenko.2015}. Assuming isotropy and a constant diffusivity $D$, the temporal evolution of the exciton density is given by 
\begin{align} \label{modelling_ex_diff_eq}
	\frac{\partial n_\mathrm{LE}}{\partial t} = D \Delta n_\mathrm{LE} + G_\mathrm{LE} - R_\mathrm{LE} \quad \mathrm{on}~\Omega
\end{align}  
where $\Delta$ is the Laplace operator. The first term on the right side of \eqref{modelling_ex_diff_eq} describes diffusion, the second generation (e.g.\ by illumination) and the third recombination (e.g.\ monomolecular decay) of excitons. \\
Charge generation in organic solar cells is believed to proceed via an intermediate CT state, which may form when an exciton reaches the donor-acceptor interface \cite{Vandewal.2016}. As the CT state is located at the donor-acceptor interface, the number of sites where CT excitons can be formed scales with the interfacial area. Therefore, it seems natural to define the CT density $n_\mathrm{CT}(\mathbf{x},t)$ as a surface density (so that the total number of CT excitons can be calculated by the integral over $\Gamma_\mathrm{d}$, see \eqref{modelling_conservation}). To describe the charge generation succeeding the diffusion process, we use a standard rate model, displayed in figure \textbf{\ref{fig:modelling}b)}, which incorporates all relevant species and processes. Excitons can form the CT state when they are in the vicinity of the interface. In our model, however, excitons dissociate only directly at the interface. This is illustrated in figure \textbf{\ref{fig:modelling}b)} by introducing the exciton density at the interface $n_\mathrm{LE}^{\ast}$. Therefore, following the work of Riley \etal \cite{Riley.2022}, we will account for exciton dissociation and reformation to and from the CT state by invoking a Robin-type boundary condition for \eqref{modelling_ex_diff_eq}
\begin{align} \label{modelling_diss_boundary}
	- D \nabla n_\mathrm{LE} \cdot \boldsymbol{\nu} = \kappa_\mathrm{diss,LE} n_\mathrm{LE} - k_\mathrm{ref} n_\mathrm{CT} \quad \mathrm{on}~\Gamma_\mathrm{d}
\end{align}
where $\nabla$ is the nabla operator, $\kappa_\mathrm{diss,LE}$ is the exciton dissociation constant and $k_\mathrm{ref}$ is the exciton reformation rate constant. Checking the units reveals that $\kappa_\mathrm{diss,LE}$ has the unit of a velocity, while $k_\mathrm{ref}$ is a \enquote{real} rate (unit one over time), since $n_\mathrm{CT}$ is a surface density. We will argue in the following that this exciton dissociation velocity can be written as the product of a characteristic length $d$ and a \enquote{real} rate constant $k_\mathrm{diss,LE}$, $\kappa_\mathrm{diss,LE} = k_\mathrm{diss,LE} d$. Collins and Kimball dealt with this issue in their seminal work on semi-absorbing boundaries. They have demonstrated that a boundary condition of the form \eqref{modelling_diss_boundary} can be derived as a limit of a random walk, where $d$ is taken to be half the mean jump length \cite{FrankCCollins.1949}\cite{Goodrich.1954}. We give a different motivation for $d$ as the thickness of the CT localisation zone in the Supplementary Material (\textbf{\ref{sup_kappa_dissex}}). This is consistent with the results presented by de Falco $et~al.$, who performed a micro-to-macro scale transition and also arrived at a model including a CT surface density and an exciton dissociation constant $\kappa_\mathrm{diss,LE}$ that is proportional to the thickness of the CT localisation zone \cite{Falco.2012}. In both interpretations, $d$ is of the same order and is set to one nanometre in the following. \\
The reflective boundary condition is simply
\begin{align} \label{modelling_reflective_boundary}
	\nabla n_\mathrm{LE} \cdot \boldsymbol{\nu} = 0 \quad \mathrm{on}~\Gamma_\mathrm{r}.
\end{align}
The temporal evolution of the CT density is similar as in standard rate equations \cite{Sandberg.2021} 
\begin{align} \label{modelling_CT}
	\frac{\partial n_\mathrm{CT}}{\partial t} = \kappa_\mathrm{diss,LE} n_\mathrm{LE} - (k_\mathrm{ref} + k_\mathrm{diss,CT})n_\mathrm{CT} + G_\mathrm{CT} - R_\mathrm{CT} \quad \mathrm{on}~\Gamma_\mathrm{d}.
\end{align}
Again, the last two terms denote generation and (geminate) recombination. $k_\mathrm{diss,CT}$ is the rate constant describing the dissociation of CT excitons to free charges. Consequently, the generation rate of free carriers reads
\begin{align} \label{modelling_CS}
	\frac{dN_\mathrm{CS}}{dt} = \int_{\Gamma_\mathrm{d}} k_\mathrm{diss,CT} n_\mathrm{CT} dA
\end{align}
where $N_\mathrm{CS}$ is the total number of charges generated by excitons initially in the sample domain. We neglect non-geminate recombination of charges as we are only interested in free charge generation. Further, we set the CT concentration and the charge population prior to illumination to zero by introducing the following initial conditions
\begin{align} \label{modelling_initial_cond}
	n_\mathrm{LE}(\mathbf{x},t=0) = n_{\mathrm{LE},0}(\mathbf{x}) \quad \mathbf{x} \in \Omega, \quad n_\mathrm{CT}(\mathbf{x},t=0) = 0 \quad \mathbf{x} \in \Gamma_\mathrm{d}, \quad N_\mathrm{CS}(t=0) = 0.
\end{align} 
In total, equations \eqref{modelling_ex_diff_eq} - \eqref{modelling_initial_cond} state a well-defined problem. The system thus defined explicitly ensures that excitons are only generated and eliminated by their respective generation and recombination processes
\begin{align} \label{modelling_conservation}
	&\frac{d}{dt} \left[\int_\Omega n_\mathrm{LE} dV + \int_{\Gamma_\mathrm{d}} n_\mathrm{CT} dA + N_\mathrm{CS} \right] \notag \\
	&= \int_\Omega \frac{\partial n_\mathrm{LE}}{\partial t} dV + \int_{\Gamma_\mathrm{d}} \frac{\partial n_\mathrm{LE}}{\partial t} dA + \frac{d N_\mathrm{CS}}{dt} \notag \\
	&= \int_\Omega (G_\mathrm{LE}-R_\mathrm{LE}) dV + \int_\Gamma D \nabla n_\mathrm{LE} \cdot \boldsymbol{\nu} dA + \int_{\Gamma_\mathrm{d}} (\kappa_\mathrm{diss,LE} n_\mathrm{LE} - k_\mathrm{ref}n_\mathrm{CT} + G_\mathrm{CT} - R_\mathrm{CT}) dA \notag \\
	&= \int_\Omega (G_\mathrm{LE}-R_\mathrm{LE}) dV + \int_{\Gamma_\mathrm{d}} (G_\mathrm{CT} - R_\mathrm{CT}) dA . 
\end{align}
From the second to the third line we have used equations \eqref{modelling_ex_diff_eq}, \eqref{modelling_CT} and \eqref{modelling_CS} and the divergence theorem, while we applied the boundary conditions \eqref{modelling_diss_boundary} and \eqref{modelling_reflective_boundary} in the last step. \\
In the remainder of this paper we will neglect direct generation into the CT state ($G_\mathrm{CT}=0$), which is a reasonable assumption under one sun illumination owing to its comparatively low absorption coefficient \cite{Sandberg.2021}. The decay of CT excitons to the ground state (geminate recombination) will be expressed as
\begin{align} \label{modelling_CT_decay}
	R_\mathrm{CT} = k_\mathrm{f,CT} n_\mathrm{CT}.
\end{align}
$k_\mathrm{f,CT}$ is the sum of the radiative and non-radiative CT decay rates. Moreover, all the introduced rates are set constant in space and time. \\
In the following we will consider two exemplary geometries to describe specific OSC designs.
\subsection*{Spherical Geometry}
The acceptor domains in BHJ-based OSCs employing NFAs are usually approximated to be round \cite{Sajjad.2020} and a spherical geometry has been utilized to study exciton diffusion and dissociation before \cite{Whaley.2014}\cite{Hedley.2013}. We stick to this domain shape approximation and consider $\Omega$ to be a sphere with radius $R$ as shown in figure \textbf{\ref{fig:modelling}c)}. For simplicity, in this BHJ model we assume that the spherical acceptor grain is completely encompassed by the donor phase, i.e.\ $\Gamma = \Gamma_\mathrm{d}$. Additionally, we will assume $R$ to be sufficiently small to neglect changes in the light intensity across the domain. This implies that the exciton generation (either $G_\mathrm{LE}$ or $ n_{\mathrm{LE},0}$, see below) is uniform within the domain and therefore spherically symmetric. Together with standard exciton recombination processes (see equation \eqref{exp_full_recomb}) this leads to a spherical symmetric solution $n_\mathrm{LE}(r,t)$ and a quasi one dimensional diffusion problem \cite{Crank.1979}. The variable $r$ denotes the distance from the centre of the sphere. Furthermore, the spherical symmetry implies that the CT density will be uniform across the entire surface, $n_\mathrm{CT} = n_\mathrm{CT}(t)$. Under these assumptions the spherical model can be outlined as 
\begin{align}
    &\frac{\partial n_{\mathrm{LE}}}{\partial t}(r,t) = \frac{1}{r^2}\frac{\partial}{\partial r}\left(r^2\frac{\partial n_{\mathrm{LE}}}{\partial r}(r,t)\right) + G_\mathrm{LE}(t) - R_\mathrm{LE}(r,t), \quad 0 \leq r \leq R \label{modelling_sphere_diffu} \\
    &\frac{\partial n_{\mathrm{LE}}}{\partial r}(0,t) = 0, \quad -D \frac{\partial n_\mathrm{LE}}{\partial r}(R,t) = \kappa_\mathrm{diss,LE} n_\mathrm{LE}(R,t) - k_\mathrm{ref} n_\mathrm{CT}(t) \label{modelling_sphere_boundary}\\
    &\frac{d n_\mathrm{CT}}{d t}(t) = \kappa_\mathrm{diss,LE} n_\mathrm{LE}(R,t) - (k_\mathrm{ref} + k_\mathrm{f,CT}+k_\mathrm{diss,CT})n_\mathrm{CT}(t) \label{modelling_sphere_CT}\\
    &\frac{dN_\mathrm{CS}}{dt}(t) = k_\mathrm{diss,CT}4 \pi R^2 n_\mathrm{CT}(t) \label{modelling_sphere_CGrate} \\
    &n_\mathrm{LE}(r,0) = n_{\mathrm{LE},0}, \quad n_\mathrm{CT}(0) = 0, \quad N_\mathrm{CS}(0) = 0. \label{modelling_sphere_IC}
\end{align}
Note that the first equation in \eqref{modelling_sphere_boundary} is a necessary symmetry condition \cite{Langtangen.2017}.
Equation \eqref{modelling_sphere_CT} is solved by 
\begin{align} 
	&n_\mathrm{CT}(t) = \kappa_\mathrm{diss,LE} \int_0^t \exp(k(s-t)) n_\mathrm{LE}(R,s) ds, \label{modelling_sphere_CT_sol} \\
	&k \coloneq k_\mathrm{ref} + k_\mathrm{f,CT}+k_\mathrm{diss,CT}
\end{align} 
with the unknown function $n_\mathrm{LE}(R,t)$. The solution can be easily verified by plugging \eqref{modelling_sphere_CT_sol} back into the differential equation. Inserting \eqref{modelling_sphere_CT_sol} into the second equation of \eqref{modelling_sphere_boundary} formally decouples the system. This, however, leads to an integro-differential system for $n_\mathrm{LE}$, which is hard to deal with \cite{Filipov.2023} even when considering only linear exciton recombination processes. Therefore, we are using a simple method of lines (MOL) approach together with central finite differences to solve the system at hand (for details see Supplementary Note \textbf{\ref{sup_numerics}}).
\subsection*{Planar Geometry}
To describe a bilayer OSC, we use the planar geometry shown in figure \textbf{\ref{fig:modelling}d)}. A cuboidal acceptor layer with length $L$ and surface area $A$ deposited on glass (dark grey area) is covered by a donor layer (red area). The distance from the glass interface is denoted by $x$. In the case of light incidence perpendicular to the coloured surfaces, $G_\mathrm{LE}$ and $n_\mathrm{LE,0}$ are only dependent on $x$ in accordance with the Beer-Lambert law. By the same reasoning as above, the diffusion problem becomes one dimensional in space $n_\mathrm{LE} = n_\mathrm{LE}(x,t)$ and the CT density depends solely on time. The plain $x=0$ defines a reflecting boundary while the interface $x=L$ enables exciton dissociation. The resulting equations are similar to the ones of the spherical system and are given in the Supplementary Material (\textbf{\ref{sup_planar}}). The same numerical scheme as above is used. 

\section{Steady State Solution and Charge Generation Efficiency}
\begin{figure}[t]
	\includegraphics[width = 1 \textwidth]{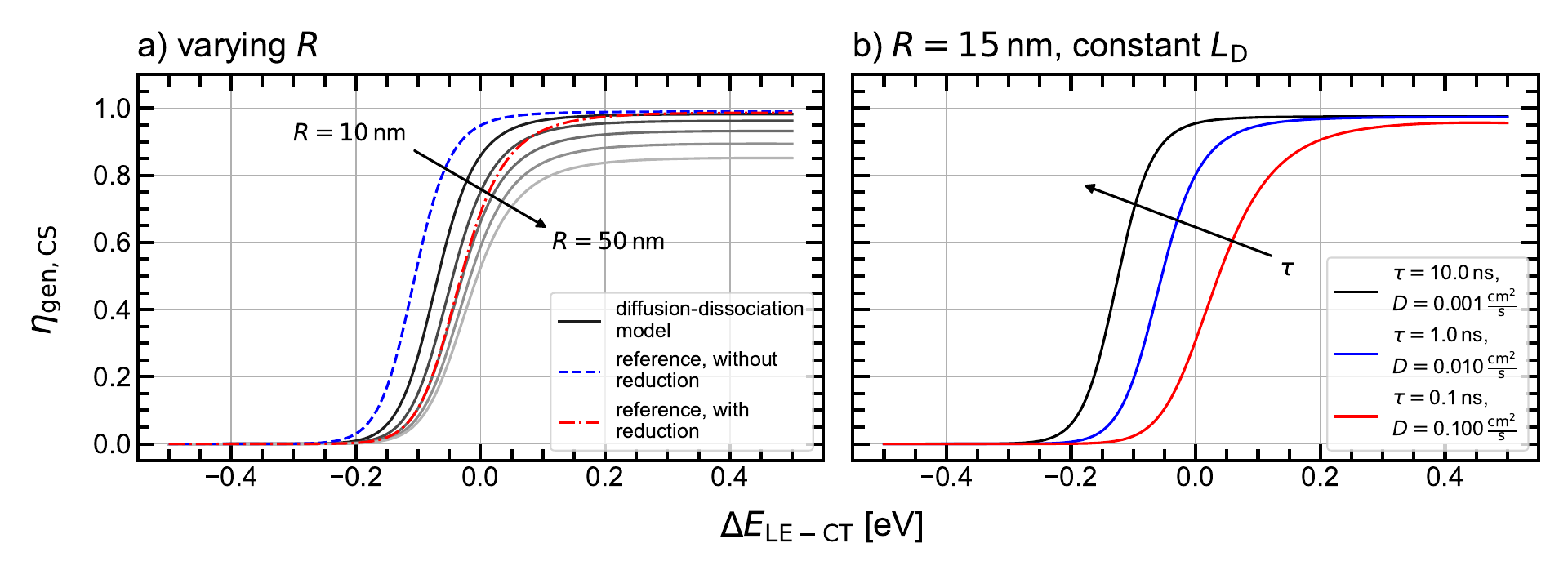}
	\caption{The charge generation efficiency $\eta_\mathrm{gen,CS}$ as a function of the energetic offset between the local exciton and the interfacial CT state $\Delta E_\mathrm{LE-CT}$ using Marcus theory. $k_\mathrm{f,CT}=\frac{1}{1000}\,\mathrm{ps}^{-1}$, $k_\mathrm{diss,CT} = 0.1\,\mathrm{ps}^{-1}$, $d=1\,$nm, $\lambda=0.45\,$eV, $|H_\mathrm{DA}| = 0.01\,$eV and $T=293.15\,$K. \textbf{a)} The domain size radius $R$ is varied from $10\,$nm (dark) to $50\,$nm (light) in steps of $10\,$nm while the diffusion length $L_\mathrm{D} \approx 31.6\,$nm  ($k_\mathrm{f,LE}=\frac{1}{1000}\,\mathrm{ps}^{-1}$, $D=10^{-2}\,\frac{\mathrm{cm}^2}{\mathrm{s}}$) is kept constant. As a reference, the rate model (see Supplementary Note \textbf{\ref{sup_ref_model}}) without accounting for diffusion explicitly is plotted with and without the reduction factor of $0.1$ \cite{Pranav.2024}. \textbf{b)} The intrinsic lifetime $\tau$ is demonstrated to be significant for charge generation in OSCs. To this end, $\tau$ is increased while the diffusivity $D$ is decreased so that the diffusion length remains constant at the same value as in a). This improves the charge generation efficiency, especially at low offsets. The domain size is fixed to $R=15\,$nm.}
	\label{fig:steady_state_efficiencies}
	\centering
\end{figure}

To investigate the impact of a varying HOMO offset on charge generation in OSCs, recent studies have used a rate equation model in which the rates of singlet exciton dissociation into CT states and reformation from CT states are described by Marcus theory of electron transfer \cite{Pranav.2024}. This model was able to properly describe the efficiency of charge generation as a function of the energetic offset at the donor-acceptor heterojunction for a wide range of donor:NFA combinations. To take into account that only a fraction of the singlet excitons are situated at the donor-acceptor interface and able to form a CT state, it is common practice to weight the exciton dissociation rate $k_\mathrm{diss,LE}$ with a prefactor smaller than one (typically $0.1$). This implicitly includes exciton diffusion. However, it remains unclear how the prefactor relates to the exciton diffusivity or domain size. In previous work, Riley \etal addressed this issue by explicitly including exciton diffusion into their expression for the charge generation efficiency \cite{Riley.2022}. Nevertheless, this work does not account for exciton diffusion in a three-dimensional geometry nor evaluate the critical role of the HOMO offset via Marcus theory. The latter has proved to describe hole transfer correctly for a large class of donor:NFA blends \cite{Zhou.2021}. In this section we use the spherical model developed in section \textbf{\ref{sec:modelling}} in conjunction with Marcus theory to examine the charge generation in BHJs as a function of the energetic offset for HT. Recent studies have demonstrated that hole transfer from the lower bandgap NFA to the donor is the dominant charge generation channel in NFA-based OSCs \cite{Karuthedath.2021}. In the case of donor excitation, charge generation proceeds via energy transfer from donor to NFA, followed by HT. Hence, it is sufficient to consider HT as the dominant charge generation mechanism in NFA-based OSCs. \\
We consider the situation of a consistently illuminated OSC, wherein exciton generation is time-independent, $G_\mathrm{LE} = G = \mathrm{const}$, and a quasi steady-state for singlet and CT excitons is reached. To model this steady-state (here an initial condition is not needed), we set the time derivatives of equations \eqref{modelling_sphere_diffu} and \eqref{modelling_sphere_CT} to zero. For the sake of simplicity, we will denote the steady-state solutions throughout this section by $n_\mathrm{LE}$ and $n_\mathrm{CT}$, too. At one sun illumination, the exciton concentration in the bulk is sufficiently small so that we can neglect exciton-exciton interactions such as singlet-singlet annihilation (SSA) \cite{Kohler.2015}. Therefore, monomolecular decay is the dominant singlet exciton recombination mechanism
\begin{align} \label{steady_recomb_low_fluence}
	R_\mathrm{LE} = k_\mathrm{f,LE} n_\mathrm{LE}
\end{align}
where $k_\mathrm{f,LE}$ is again the sum of the radiative and non-radiative singlet exciton decay rates. The inverse of the total decay rate is the intrinsic exciton lifetime $\tau = k_\mathrm{f,LE}^{-1}$. The decay of excitons in CT state can occur via geminate recombination as given in equation \eqref{modelling_CT_decay}. For the specified generation and recombination processes the steady-state solution can be calculated analytically
\begin{align} 
	&n_{\mathrm{LE}}(r) = \frac{G}{k_\mathrm{f,LE}}\left(1-\frac{\kappa_\mathrm{diss,LE}(1-P_\mathrm{ref})R}{k_\mathrm{f,LE}L_\mathrm{D}r} \frac{\sinh{\left(\frac{r}{L_\mathrm{D}}\right)}}{\cosh{\left(\frac{R}{L_\mathrm{D}}\right)}\left[\tanh{\left(\frac{R}{L_\mathrm{D}}\right)}\left[\frac{\kappa_\mathrm{diss,LE}(1-P_\mathrm{ref})}{k_\mathrm{f,LE}L_\mathrm{D}}-\frac{L_\mathrm{D}}{R}\right]+1\right]} \right) \label{steady_nLE} \\
	&n_\mathrm{CT} = \frac{\kappa_\mathrm{diss,LE}}{k_\mathrm{ref} + k_\mathrm{f,CT}+k_\mathrm{diss,CT}} n_\mathrm{LE}(R) \label{steady_nCT} \\
	&L_\mathrm{D} \coloneq \sqrt{\frac{D}{k_\mathrm{f,LE}}}, \quad P_\mathrm{ref} \coloneq \frac{k_\mathrm{ref}}{k_\mathrm{ref}+k_\mathrm{f,CT}+k_\mathrm{diss,CT}} 
\end{align}
where we have introduced the diffusion length $L_\mathrm{D}$ and the probability $P_\mathrm{ref}$ of reforming LE from CT. The mean squared displacement (MSD) of an exciton during its intrinsic lifetime is $\mathrm{MSD}=6D\tau$ (given that exciton does not reach the donor-acceptor interface during that time), since the exciton diffuses in three dimensions. This would lead to a diffusion length of $\sqrt{6D\tau}$. However, we adhere to the above definition, as it is more convenient for our equations. \\
The charge generation efficiency $\eta_\mathrm{gen,CS}$ can now be calculated by dividing \eqref{modelling_sphere_CGrate} by the total number of generated excitons per unit time
\begin{align}
	\eta_\mathrm{gen,CS} &= \frac{k_\mathrm{diss,CT}4 \pi R^2 n_\mathrm{CT}}{\frac{4}{3}\pi R^3 G} = \frac{4 \pi R^2 \kappa_\mathrm{diss,LE} n_\mathrm{LE}(R) P_\mathrm{diss,CT}}{\frac{4}{3}\pi R^3 G} \notag \\
	&= 3\frac{\kappa_\mathrm{diss,LE}}{R k_\mathrm{f,LE}} P_\mathrm{diss,CT} \frac{1}{1+\frac{\frac{\kappa_\mathrm{diss,LE}}{R k_\mathrm{f,LE}}(1-P_\mathrm{ref})\frac{R}{L_\mathrm{D}}\tanh{\left(\frac{R}{L_\mathrm{D}}\right)}}{1-\frac{L_\mathrm{D}}{R}\tanh{\left(\frac{R}{L_\mathrm{D}}\right)}}} \label{steady_CGeff} \\
	P_\mathrm{diss,CT} &\coloneq \frac{k_\mathrm{diss,CT}}{k_\mathrm{ref}+k_\mathrm{f,CT}+k_\mathrm{diss,CT}}.
\end{align}
$P_\mathrm{diss,CT}$ is the probability of CT excitons forming separated charges before reforming to LE or decaying to the ground state. $\eta_\mathrm{gen,CS}$ can be expressed exclusively by the following four combinations of parameters: a characteristic length ratio $l \coloneq \frac{L_\mathrm{D}}{R}$, a characteristic velocity ratio $v \coloneq \frac{\kappa_\mathrm{diss,LE}}{R k_\mathrm{f,LE}} = \frac{k_\mathrm{diss,LE}d}{R k_\mathrm{f,LE}}$, as well as $(1-P_\mathrm{ref})$ and $P_\mathrm{diss,CT}$. When $k_\mathrm{f,CT} \ll k_\mathrm{diss,CT},$ as it is usually the case in most of the efficient NFA-based OSCs \cite{Tamai.2017}\cite{Natsuda.2022}, we have $1-P_\mathrm{ref} \approx P_\mathrm{diss,CT}$. In this case the charge generation efficiency just depends on the two parameters $l$ and $vP_\mathrm{diss,CT}$. Increasing $l$ and $v$ always leads to an increase in the charge generation efficiency, which can be checked by calculating the respective partial derivatives. In the limit of $l$ and $v$ tending to infinity, $\eta_\mathrm{gen,CS}$ is just determined by the dissociation of the CT state $\lim_{l \to \infty} \lim_{v \to \infty} \eta_\mathrm{gen,CS}=\frac{P_\mathrm{diss,CT}}{1-P_\mathrm{ref}}=\frac{k_\mathrm{diss,CT}}{k_\mathrm{diss,CT}+k_\mathrm{f,CT}}$ as one would expect. Similar to the charge generation efficiency, the fraction of generated excitons that decay via CT, $\eta_\mathrm{f,CT}$, or in the bulk, $\eta_\mathrm{f,LE}$, can be determined (see Supplementary Note \textbf{\ref{sup_decay_eff}}). As expected, those three quantities add up to one. \\
To calculate the exciton dissociation and reformation rates as a function of the energetic offset between the local exciton and the interfacial CT state $\Delta E_\mathrm{LE-CT} \coloneq E_\mathrm{LE} - E_\mathrm{CT}$ (see figure \textbf{\ref{fig:modelling}a)}) we use Marcus theory \cite{Marcus.1985} in the semi-classical limit
\begin{align} \label{steady_Marcus}
	k_\mathrm{diss,LE/ref} &= \frac{|H_\mathrm{DA}|^2}{\hbar} \sqrt{\frac{\pi}{\lambda k_\mathrm{B}T}} \exp{\left(-\frac{(\lambda \mp\Delta E_\mathrm{LE-CT})^2}{4 \lambda k_\mathrm{B}T}\right)}
\end{align}
where $-$ corresponds to dissociation and $+$ to reformation. In equation \eqref{steady_Marcus}, $|H_\mathrm{DA}|$ represents the electronic matrix element describing the donor-acceptor coupling for HT, $\lambda$ the corresponding reorganisation energy and $T$ the temperature. It should be noted that Pranav \etal employed the energy difference $\Delta E_\mathrm{CT-LE} = -\Delta E_\mathrm{LE-CT}$ instead of $\Delta E_\mathrm{LE-CT}$ to investigate the free charge generation efficiency as a function of the driving force. Therefore, the appearance of the following schemes differs from those presented in \cite{Pranav.2024}. \\
In figure \textbf{\ref{fig:steady_state_efficiencies}a)} the charge generation efficiency versus the energetic offset is shown.
The used parameters are oriented towards experimental values of Y6 (see experimental section) and are summarised in the caption. We adopt the value of the electronic coupling $|H_\mathrm{DA}| = 0.01\,$eV obtained from calculations on PM6:Y6 \Cite{Pranav.2024} while we choose $\lambda = 0.45\,$eV. This can be regarded as some intermediate value between reported low and high reorganisation energies of $\sim 0.33\,$eV \cite{Zhong.2020} and $\sim 0.55\,$eV \cite{Pan.2017} respectively for typical polymer:NFA blends. To investigate the influence of diffusion, we varied the domain radius between $10\,$nm ($l \approx 3.16$) and $50\,$nm ($l \approx 0.63$). This results in a decrease of both $l$ and $v$, which in turn reduces the efficiency. As a reference, we have plotted the efficiency obtained from the standard rate model introduced by Pranav \etal \cite{Pranav.2024} (see Supplementary Note \textbf{\ref{sup_ref_model}}) with and without the reduction factor of $0.1$ in the exciton dissociation rate. The reduction factor is typically introduced because only excitons near the interface can dissociate immediately after being generated. The majority of excitons are located in the bulk and must undergo a process of diffusion before undergoing dissociation at the interface. In contrast, the singlet exciton reformation rate is not reduced since every CT state is surrounded by at least one neutral molecule \cite{Pranav.2024}. \\
When $R$ is decreased towards zero, $\eta_\mathrm{gen,CS}$ approaches the reference curve without a reduction factor. This makes sense as diffusion plays only a minor role in small domains. When the reduction factor is included, the reference falls nicely into the family of curves representing reasonable domain sizes. However, the reference curve does not coincide exactly with the efficiency curve of one specific domain size. Furthermore, both reference efficiencies tend to unity (or more generally to $\frac{k_\mathrm{diss,CT}}{k_\mathrm{diss,CT}+k_\mathrm{f,CT}}$) for high energetic offsets, meaning that the energy of the CT state is well below LE. In contrast, our model efficiencies at large domain sizes do not reach the maximum efficiency of one despite an offset equal to the reorganisation energy. In that case, charge generation is limited by excitons decaying in the bulk before reaching the donor-acceptor interface by diffusion. This demonstrates that the complex interplay between exciton diffusion and dissociation cannot be fully captured by the conventional prefactor in standard rate equation models. \\
As the offset approaches zero, a critical offset point is reached wherein the generation efficiency begins to drop significantly due to the slowing of exciton dissociation and the increased probability of reformation. Thereby, the majority of excitons decay while still being in the bulk (see Supplementary Figure \textbf{\ref{fig:sup_efficiencies}}). These results align with experimental observations on a wide range of donor:NFA combinations \cite{Zhou.2021}\cite{Pranav.2024}. The latter was demonstrated via the anticorrelation between the field-dependence of free charge generation and photoluminescence \cite{Pranav.2023}. The critical offset point at which the lack of driving force begins to compromise charge generation is shifted to higher offsets when the domain radius is increased as a result of the longer exciton diffusion time. We will discuss later that the average exciton diffusion time scales with $R^2$ (see \eqref{times_MFPT}). Interestingly, this drop of the charge generation efficiency aligns with the pronounced reduction in non-radiative $V_\mathrm{OC}$ losses when the offset is reduced to small values, which has been reported in \cite{Classen.2020} for NFA-based OSCs. Since both effective charge generation and low non-radiative $V_\mathrm{OC}$ losses are important to maximise the power conversion efficiency, this emphasises the necessity to achieve near unity charge generation at negligible offsets. \\
As discussed before, decreasing the domain size enables efficient charge generation at small driving forces. At the same time, well-mixed donor and acceptor layers hinder the  extraction of free charges due to the large extent of interfaces and the associated recombination losses \cite{Zhang.2019}. Therefore, the diffusion length $L_\mathrm{D}$ must be optimised to achieve high characteristic length ratios $l$. This can be done by either increasing the diffusivity $D$ or the intrinsic lifetime $\tau$. As has been pointed out by Riley $et~al.$, the pivotal parameter is the latter \cite{Riley.2022}. Our model confirms this conclusion, as illustrated in figure \textbf{\ref{fig:steady_state_efficiencies}b)}, where the efficiency is plotted again as a function of the energetic offset. We fix the radius to $15\,$nm and vary the ratio of $\tau$ and $D$ while keeping the diffusion length constant. For energetic offsets between $-0.2\,$eV and $0.2\,$eV, the combination of a high intrinsic lifetime and a low diffusion constant results in a notable enhancement in efficiency. While both the intrinsic lifetime and the diffusion constant influence the probability of an exciton reaching the interface due to their similar effect on the diffusion length, a high intrinsic lifetime also allows for multiple dissociation attempts when an exciton reaches the interface initially or when it is reformed from the CT state. This is also reflected in the characteristic constants, as the lifetime also affects the velocity ratio $v$, whereas the diffusivity exerts no influence on this quantity. In our illustrative example, an exceptionally long exciton lifetime of $10\,$ns can facilitate charge generation efficiencies close to unity in spite of a vanishing driving force and a moderate diffusion length. The positive effect of prolonged lifetimes has also been observed experimentally \cite{Classen.2020} and should be considered in the future development of new OSC materials.

\section{Characteristic Times of Charge Generation}
\begin{figure}[htpb!]
	\includegraphics[width = 1 \textwidth]{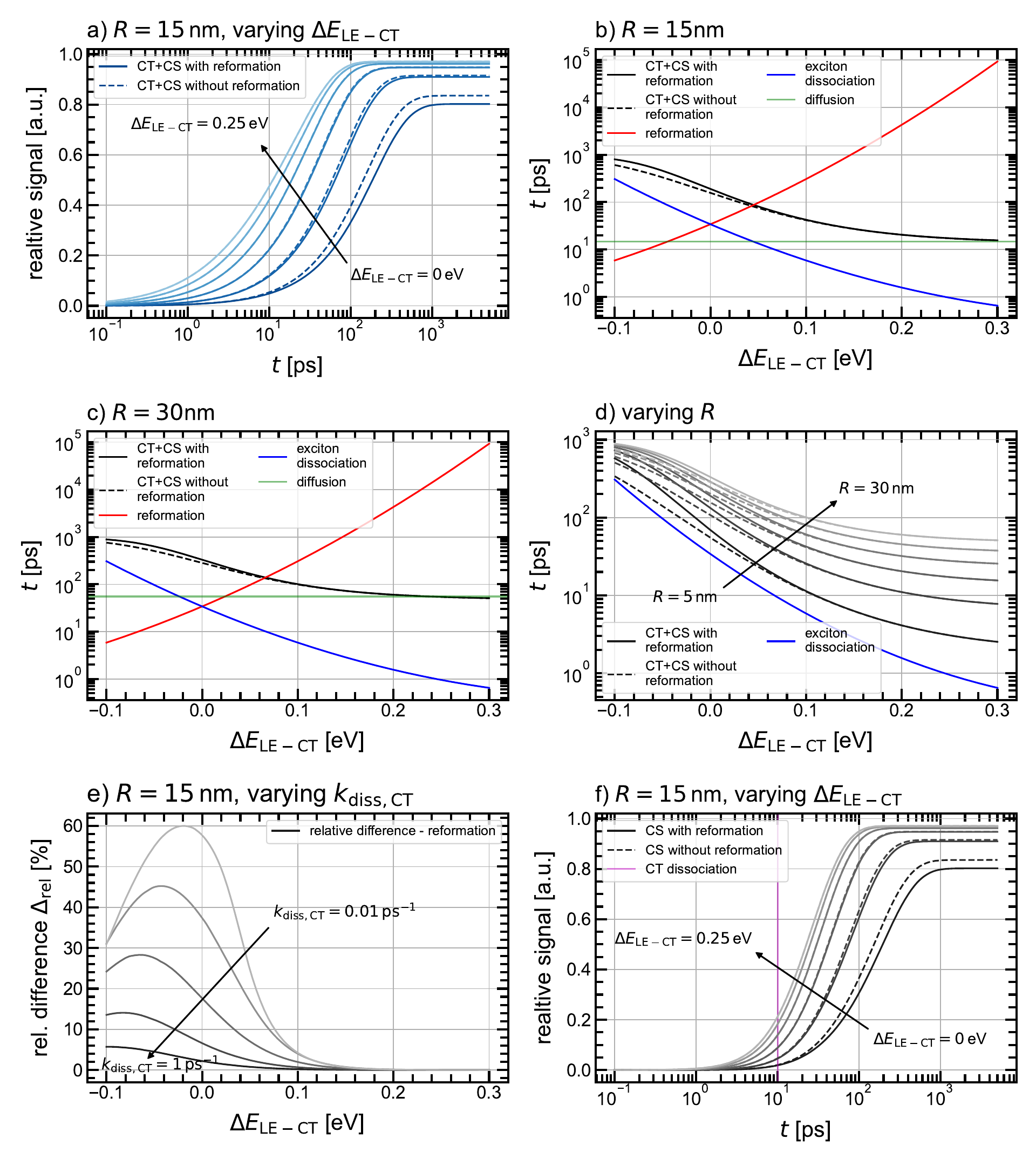}
	\caption{$k_\mathrm{f,LE}=k_\mathrm{f,CT}=\frac{1}{1000}\,\mathrm{ps}^{-1}$, $k_\mathrm{diss,CT} = 0.1\,\mathrm{ps}^{-1}$, $D=10^{-2}\,\mathrm{cm}^2\mathrm{s}^{-1}$, $d=1\,$nm, $\lambda=0.45\,$eV, $|H_\mathrm{DA}| = 0.01\,$eV and $T=293.15\,$K. \textbf{a)} The temporal evolution of the sum of CT and CS normalised to the number of initially generated excitons for different driving forces in a TAS-like simulation. The energetic offset $\Delta E_\mathrm{LE-CT}$ is varied from $0\,$eV (dark) to $0.25\,$eV (light) in steps of $0.05\,$eV. The respective solid line shows the full model while the dashed line displays the simulation result when reformation of excitons is not included. \textbf{b)} and \textbf{c)} Characteristic times of the GSB rise (CT+CS) and processes associated with charge generation as a function of $\Delta E_\mathrm{LE-CT}$. The GSB rise times with and without reformation are shown for two different domain sizes. Furthermore, the characteristic times of singlet exciton reformation and dissociation are indicated. The diffusion time is the MFPT of a photo-generated exciton with respect to the interface reduced by the finite exciton lifetime (see \textbf{\ref{sup_mean_times}}). \textbf{d)} The domain size is enlarged from $5\,$ nm (dark) to $30\,$nm (light) in steps of $5\,$nm. \textbf{e)} Relative difference between the characteristic GSB rise time with and without reformation. The CT dissociation rate $k_\mathrm{diss,CT}$ is changed evenly on a log scale from $0.01\,\mathrm{ps}^{-1}$ (light) to $1\,\mathrm{ps}^{-1}$ (dark). \textbf{f)} Normalised population of separated charges, proportional to the EA signal in TAS. The CT dissociation time scale of $10\,$ps is indicated by the coloured vertical line. The offset is altered as in a).}
	\label{fig:times}
	\centering
\end{figure}
Various processes affect the charge generation in OSCs, including exciton diffusion to the interface, CT state formation at the donor-acceptor interface, reformation of excitons (in low-offset systems) and dissociation of the CT state to separated charges. The entirety of those mechanisms and their time scales can be monitored by TAS \cite{Tamai.2024}. Nevertheless, there is an intricate interplay between the different processes that makes the interpretation of TAS data a challenging endeavour. The objective of this section is to disentangle the observed rise times of characteristic spectral features in TAS with respect to the various processes involved. Throughout this section, we exclude SSA from our analysis, which normally cannot be ignored when interpreting TAS data, as measurements are usually performed at relatively high fluences to achieve a substantial signal-to-noise ratio \cite{Tamai.2024}. SSA describes the collision of two diffusing excitons, which results in the formation of a highly excited state. This state typically decays back to LE via internal conversion, while one exciton is lost \cite{Kohler.2015}. Therefore, SSA leads to a faster decay of the TAS signal when increasing the fluence. On the other hand, ignoring SSA allows us to arrive at analytical predictions of characteristic timescales. Hence, we use the linear exciton recombination rate given by \eqref{steady_recomb_low_fluence}. Although the simulation of TAS dynamics is only accurate at low excitation fluences, the qualitative outcomes are nevertheless applicable to high fluences. \\
To simulate TAS, we use our spherical BHJ model. Further, our domain is considered to be an acceptor that is selectively excited in a pump-probe experiment. The pump pulse generates excitons within the time resolution of the experiment \cite{Kurpiers.2018}. Therefore, we model exciton generation by introducing a uniform initial density $n_\mathrm{LE,0}$ at $t=0$, which represents the total number density of absorbed photons. The generation rate $G_\mathrm{LE}$ is set to zero. The exact value of $n_\mathrm{LE,0}$ is irrelevant because all processes considered in the model without SSA are linear. We remind the reader here that we neglect reformation of CT states by free charge recombination, which would be a second order process. This is again because we only consider the low fluence case where charge recombination proceeds at a much slower time scale than charge generation and separation \cite{Kurpiers.2018}. In our scenario, HT from the acceptor to the donor can be tracked by the GSB of the donor. The change in optical density due to the GSB is proportional to the sum of the CT state population ($4 \pi R^2 n_\mathrm{CT}$) and free charges $N_\mathrm{CS}$. In figure \textbf{\ref{fig:times}a)} the temporal evolution of the GSB normalised to the number of initially generated excitons is plotted for different energetic offsets and a domain radius of $R=15\,$nm. Again, we apply Marcus theory with the same parameters as above to calculate the dissociation and reformation rate. Moreover, the dynamics are shown with and without reformation. The corresponding parameters can be taken from the caption. As anticipated, the effect of reformation becomes only prominent at small or negative offsets, where the CT state energy is located close to or above the LE energy. Reformation slows down the rise of the GSB and reduces the signal. \\
When increasing the driving force for HT at the interface, the GSB signal is not only stronger but also reaches its maximum faster. In the case of very high driving forces, the growth kinetics are determined entirely by diffusion, independent of the offset. To investigate the observed convergence of the GSB rise time and to clarify the contributions from the various processes involved, we define a characteristic GSB time. This is the time when the GSB signal reaches $1-e^{-1} \approx 63\,\%$ of its maximum value. When assuming an exponential-like growth, which is a good approximation for low offsets as can be seen in the Supplementary Material, the characteristic time corresponds to the mean time of charge generation (see \textbf{\ref{sup_mean_times}}, \textbf{\ref{fig:sup_GSB_vs_exp_15}} and \textbf{\ref{fig:sup_GSB_vs_exp_30}}). 
To estimate the diffusion time, we calculate the mean first passage time (MFPT) with respect to the interface, $\tau_\mathrm{diffusion}$, of a photo-generated exciton with diffusivity $D$. When the initial position of the exciton within the sphere (radius $R$) is uniformly distributed, the MFPT reads
\begin{align} \label{times_MFPT}
	\tau_\mathrm{diffusion} = \frac{2}{5} \cdot \frac{R^2}{6D}.
\end{align}
Note that this is the average time it takes when the particle is initially at the centre of the sphere weighted by a factor of $\frac{2}{5}$. After all, the time labelled by diffusion in figures \textbf{\ref{fig:times}b)} and \textbf{\ref{fig:times}c)} is not exactly the MFPT but rather a mean survival time which also incorporates the finite exciton lifetime (thus the time shown is slightly reduced compared to \eqref{times_MFPT}, for details see Supplementary Material \textbf{\ref{sup_mean_times}}).
Finally, we define characteristic times of exciton dissociation and reformation from the rates calculated by Marcus theory
\begin{align}
	\tau_\mathrm{diss,LE/ref} = \frac{1}{k_\mathrm{diss,LE/ref}}.
\end{align}
In figure \textbf{\ref{fig:times}b)} and \textbf{\ref{fig:times}c)} the results of our analysis are summarised for radii of $15\,$nm and $30\,$nm. The GSB rise time converges to the diffusion time upon increasing the offset as conjectured before. At high driving forces, exciton diffusion to the interface becomes the limiting factor of the charge generation time. In case of the large domain in figure \textbf{\ref{fig:times}c)}, the GSB rise time becomes even slightly shorter than the mean diffusion time because the defined GSB rise time is not exactly the mean time of charge generation, as the dynamics of the GSB at high offsets cannot be described solely by a single exponential (see \textbf{\ref{sup_mean_times}}, \textbf{\ref{fig:sup_GSB_vs_exp_15}} and \textbf{\ref{fig:sup_GSB_vs_exp_30}}). For small driving forces, ($\Delta E_\mathrm{LE-CT} < 0.2\,$eV) the charge generation is not only limited by diffusion any more. In fact, the characteristic GSB rise time is considerably longer than the pure diffusion time, even when reformation is ignored, while the intrinsic exciton dissociation time is still on the order of a few picoseconds. Thus, charge generation in low-offset OSCs is not simply the sum of diffusion and dissociation but the result of a more complex interplay, since the exciton may diffuse away from the interface after an unsuccessful dissociation attempt. This can also be understood with Fick's law of diffusion underlying equation \eqref{modelling_ex_diff_eq}, which states that the diffusive flux is proportional to the gradient of exciton concentration. Hence, ineffective exciton quenching at the interface inhibits the formation of a significant concentration gradient. Thereby, fewer excitons reach the interface, which in turn prevents the formation of a significant gradient within the domain. Figure \textbf{\ref{fig:sup_concentration_profile}} in the Supplementary Material summarises this reasoning via the time and offset-dependent singlet exciton concentration within a domain with a radius of $15\,$nm.
For several NFAs, the time to complete hole transfer to the donor upon selective excitation of the acceptor is reported to be $\sim 100\,$ps \cite{Pranav.2024}\cite{Zhong.2020}\cite{Chen.2022}. This is usually attributed to exciton diffusion. Our analysis, however, suggests that these long times are rather the result of a combined effect between diffusion and dissociation, given that the pure diffusion time is approximately one order of magnitude shorter than the observed charge generation times. \\
The described interplay between diffusion and dissociation becomes less relevant when the domain size is decreased, as the majority of excitons are generated close to the interface. This is demonstrated in figure \textbf{\ref{fig:times}d)}, thereby validating our approach. The GSB rise time approaches the intrinsic exciton dissociation time upon reduction of the domain radius from $30\,$nm to $5\,$nm. The figure also emphasises that reformation only affects charge generation for negligible values of $\Delta E_\mathrm{LE-CT}$, regardless of the domain size. This is also visible in figures \textbf{\ref{fig:times}a)} to \textbf{\ref{fig:times}c)}. The effect of reformation is interesting in this context, since the model includes the possibility of excitons diffusing away from the interface following reformation. With our choice of parameters, reformation only becomes significant at low driving forces because only then is the reformation time comparable to the exciton and CT (set to $10\,$ps) dissociation times. Otherwise, CT dissociation outcompetes reformation. We illustrate this with help of figure \textbf{\ref{fig:times}e)}. There we plot the relative difference between the GSB rise times with $\tau_\mathrm{GSB,r}$ and without $\tau_\mathrm{GSB,nr}$ reformation defined by
\begin{align}
	\Delta_\mathrm{rel} \coloneq \frac{\tau_\mathrm{GSB,r} - \tau_\mathrm{GSB,nr}}{\tau_\mathrm{GSB,r}}
\end{align}
as a function of $\Delta E_\mathrm{LE-CT}$. Note that when reducing $\Delta E_\mathrm{LE-CT}$, the relative difference starts to decrease after a certain negative driving force is reached. This is because the GSB rise time is bounded from above by the intrinsic exciton lifetime. Additionally, in figure \textbf{\ref{fig:times}e)} the dissociation constant $k_\mathrm{diss,CT}$ is increased evenly on a logarithmic scale. In this way the effect of reformation is substantially suppressed. It follows that in OSCs with a low driving force for HT the reformation of excitons from the CT state is concealed when a) the intrinsic exciton lifetime is high (as discussed in the previous section) and b) CT dissociation is efficient. A system characterised by a low charge generation yield mediated by short exciton lifetimes and poor CT dissociation has been reported by \cite{Shivhare.2022}. The result is also in line with equation \eqref{steady_CGeff}. Fast CT dissociation does not only increase $P_\mathrm{diss,CT}$ but also the probability of not reforming an exciton $1-P_\mathrm{ref}$ and thus enhances the efficiency via two characteristic parameters. \\
Charge separation from the CT state can be monitored by TAS via the dipole-like electric field generated by separated charges. The electric field causes a Stark shift in the energy levels of nearby molecules leading to an electro-absorption signal in the experiment \cite{Gelinas.2014}. We use the same framework as above to simulate the dynamics of the CS population for different energetic offsets with and without reformation. The CS population is again normalised to the number of initially generated excitons. The result is displayed in figure \textbf{\ref{fig:times}f)} for a CT dissociation time of $k_\mathrm{diss,CT}^{-1} = 10\,$ps. Due to the finite CT dissociation rate, the EA signal in \textbf{\ref{fig:times}f)} is delayed compared to the GSB in \textbf{\ref{fig:times}a)}. However, after $\sim 10\,$ps the CT dissociation depletes the CT state which then only acts as a sparsely populated intermediate state for free charge generation. The GSB signal of the donor is then mainly representative of separated charges and the EA reaches its maximum almost simultaneously with the GSB. The maximum value of the simulation corresponds to the fraction of generated excitons that are converted to free charges. This value is equal to the charge generation efficiency calculated by equation \eqref{steady_CGeff} (see Supplementary Note \textbf{\ref{sup_charge_gen_pulsed}}). The impact of reformation on this quantity is small because of efficient CT dissociation, as we have argued before. 

\section{Domain Size Estimation from TAS Dynamics}
To demonstrate that our model encompasses all relevant processes for charge generation, we compare our simulations with experimental TAS data. The signatures that we will use in our analysis are the GSB of the donor (CT+CS) and the EA signal (CS), as outlined in the preceding section. Our material system of choice is the well-researched blend PM6:Y6, as several material constants required for our model have been measured for this combination. We apply our spherical model to a BHJ and estimate the average domain size by fitting our model to the data. \\
We choose an excitation wavelength of around $800\,$nm to selectively excite Y6, e.g.\ see \cite{Wen.2021} for the absorption spectrum. As mentioned before, SSA cannot be neglected under TAS conditions. Therefore, we have included SSA in our model according to
\begin{align} \label{exp_full_recomb}
	R_\mathrm{LE} = k_\mathrm{f,LE} n_\mathrm{LE} + \frac{1}{2}k_\mathrm{SSA} n_\mathrm{LE}^2.
\end{align}
The factor of $\frac{1}{2}$ takes into account that the annihilation of a pair leaves one exciton behind \cite{Engel.2006}. The applied numeric values of the material constants are listed in table \textbf{\ref{tab:constants_Y6}}. 
\begin{table}[h!]
	\centering
	\caption{\label{tab:constants_Y6} Applied model parameters to compare with experimental TAS data of PM6:Y6.}
	\begin{tabular}{cccccccc}
		$D$ & $k_\mathrm{f,LE}$ & $k_\mathrm{SSA}$ & $|H_\mathrm{DA}|$ & $\lambda$ &  $\Delta E_\mathrm{LE-CT}$ & $d$ & $k_\mathrm{f,CT}$ \\
		\hline
		$0.017\,\frac{\mathrm{cm}^2}{\mathrm{s}}$\cite{LoGerfoM.2023} & $1\,\mathrm{ns}^{-1}$ & $ 1.36 \cdot 10^{-7}\,\frac{\mathrm{cm}^3}{\mathrm{s}}$\cite{Firdaus.2020} & $0.01\,$eV\cite{Pranav.2024} & $0.45\,$eV & $0.11\,$eV\cite{Sun.2023} & $1\,$nm & $1\,\mathrm{ns}^{-1}$
	\end{tabular}
\end{table}
Again, Marcus theory \eqref{steady_Marcus} is used to calculate the charge transfer rates at the interface. We adopt the electronic coupling calculated in \cite{Pranav.2024} while we stick to the intermediate value of $0.45\,$eV for the reorganisation energy. The intrinsic decay rate of singlet excitons $k_\mathrm{f,LE}$ is estimated to be $1\,\mathrm{ns}^{-1}$, which is within the range of reported literature values. The CT decay rate is assumed to be same as for excitons in the bulk. However, both decay constants are of less importance since other mechanisms, such as SSA and CT dissociation, occur at a much faster rate. We use the diffusion coefficient $D$ as reported by van Hulst \etal \cite{LoGerfoM.2023} using spatiotemporally resolved photoluminescence microscopy, which provides a comprehensive description of the underlying physics. To approximate the initial density $n_\mathrm{LE,0}$ in the spherical domain, we assume that photons are only absorbed by Y6 molecules and that each absorbed photon results in the creation of an exciton
\begin{align}
	n_\mathrm{LE,0} = \mathrm{Abs} \frac{F\lambda_\mathrm{ex}}{d h c} (\mathrm{ratio}+1).
\end{align}
Abs denotes the absorbed fraction of photons by the entire film, $F$ the excitation fluence, $\lambda_\mathrm{ex}$ the excitation wavelength, $d$ the film thickness and ratio the mixing ratio between PM6 and Y6 ($\frac{1}{1.2}$).
\begin{figure}[t]
	\includegraphics[width = 1 \textwidth]{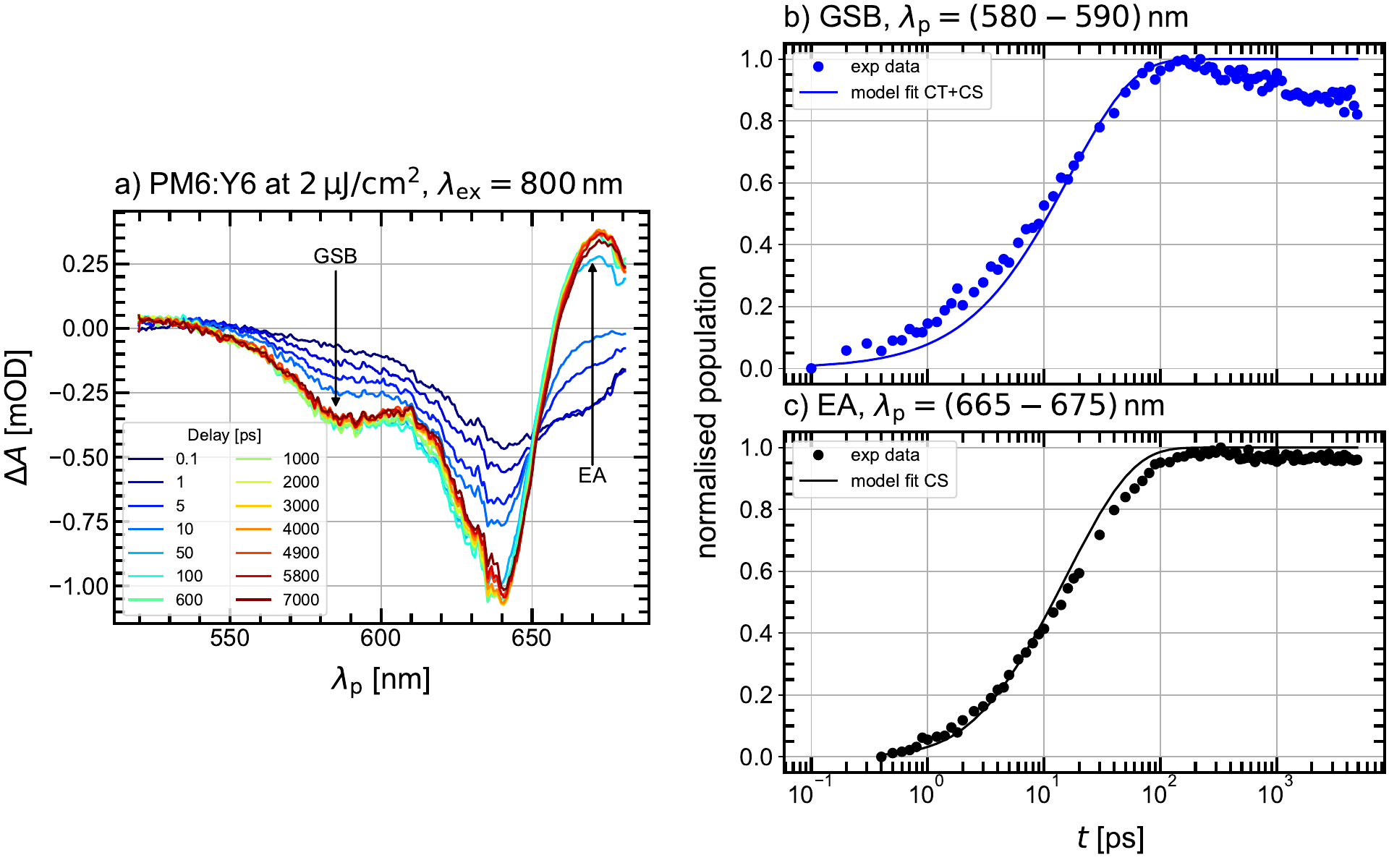}
	\caption{TAS spectra and kinetics of a BHJ PM6:Y6 blend. \textbf{a)} Smoothed spectrum for different delay times. The GSB of PM6 and the EA signal are indicated by the respective arrows. \textbf{b)} The kinetics of the GSB of PM6 obtained by averaging the signal between $\lambda_\mathrm{p}=580\,$nm and $\lambda_\mathrm{p}=590\,$nm. The initial offset ($\approx 20\%$ of the maximum signal) has been subtracted and the resulting data have been normalised to their maximum value. \textbf{c)} The kinetics of the EA signal in the wavelength range $\lambda_\mathrm{p}=665\,$nm to $\lambda_\mathrm{p}=675\,$nm. Only delay times exceeding $0.3\,$ps have been included and the data have been normalised to their maximum. The kinetics have been fitted together with the GSB in b), with the domain radius $R$ and the CT dissociation rate $k_\mathrm{diss,CT}$ as fit parameters.}
	\label{fig:exp_BHJ}
	\centering
\end{figure}
\\
Figure \textbf{\ref{fig:exp_BHJ}a)} displays the measured TAS spectrum of the BHJ as a function of the probe wavelength $\lambda_\mathrm{p}$ for different delay times. The GSB of PM6 around $\lambda_\mathrm{p} = 580\,$nm is visible initially, indicating ultrafast HT. Yet, Marcus theory predicts an exciton dissociation rate of only $k_\mathrm{diss,LE} \approx 0.2\,\mathrm{ps}^{-1}$ that does not agree with the observed ultrafast component. We attribute the initial signal to excitons generated in amorphous and more disordered Y6 domains. Those non aggregated Y6 molecules are characterised by a lower lying HOMO level, leading to a higher driving force for exciton dissociation, which facilitates ultrafast HT. Another reason for the initial timescale of HT being significantly faster than predicted by Marcus theory may be the formation of delocalised excitons, which serves to reduce the Coulomb attraction of the electron-hole pair \cite{Whaley.2014}\cite{Liang.2021}. EA can be observed at a wavelength of $\lambda_\mathrm{p}=670\,$nm. The signal rises until a maximum value is reached. Subsequently, the signal stays constant during the remaining time of the experiment as can be seen in figure \textbf{\ref{fig:exp_BHJ}c)}. This suggests that free charge recombination is not present in the observed time range which aligns with the assumption of our model. In contrast, the GSB signal begins to decay after a delay time of $200\,$ps. This is demonstrated in \textbf{\ref{fig:exp_BHJ}b)}, where we subtracted the initial offset and normalised the GSB to its maximum value. The same normalisation has been applied in \textbf{\ref{fig:exp_BHJ}c)}. Comparing GSB and EA indicates that CT dissociation is highly efficient since both normalised signals rise almost simultaneously. The observed decline in the GSB cannot be explained by non-geminate recombination of free charges as this would contradict the stable signal from EA. Therefore, the reduction in the GSB signal is likely a consequence of the decay of CT states in less aggregated Y6 phases. The mobility in these phases is poor, so that CT states cannot separate into free charges. Consequently, the CT state decays via geminate recombination, which accounts for the observed reduction of the GSB. It can be excluded that geminate recombination of CT excitons in aggregated domains is responsible for the GSB decay, as this process would not be able to compete with the observed fast CT dissociation. \\
We use the parameters given in table \textbf{\ref{tab:constants_Y6}} and normalise the total CT+CS and CS population of our simulation to its respective maximum value to match the experimental data. This leaves only the CT dissociation rate $k_\mathrm{diss,CT}$ and the domain radius $R$ as free parameters. We fit the model to the experimental data using a non-linear least squared fitting method \cite{Newville.2014}. The result is shown in the figures \textbf{\ref{fig:exp_BHJ}b)} and \textbf{c)} and the best fit has been obtained with 
\begin{align}
	R = 12.47 \pm 0.71\,\mathrm{nm}, \quad k_\mathrm{diss,CT} = 1.12 \pm 0.44\,\mathrm{ps}^{-1}
\end{align}
as parameters. The fast CT dissociation rate confirms what has been observed when comparing the kinetics of GSB and EA. The radius corresponds to a domain diameter of approximately $25\,$nm. This value lies well within the range of reported average domain sizes in efficient OSCs based on PM6:Y6, as measured by X-ray scattering methods \cite{Jiang.2019}\cite{Yuan.2019}. It has to be noted that the value of $\kappa_\mathrm{diss,LE}$ exerts a significant influence on the result, since there is a strong correlation between the exciton dissociation rate and the domain radius, with a positive correlation coefficient approaching one when both parameters are fitted simultaneously. This strong correlation underlines the connection between exciton diffusion and dissociation discussed before and shows that great care must be taken when estimating the domain size with this method. However, contrary to existing methods of determining the domain size based on exciton diffusion \cite{Sajjad.2020}\cite{Hedley.2013}, our approach considers the finite exciton dissociation rate at the interface. The impact of the finite $\kappa_\mathrm{diss,LE}$ is significant, even at moderate offsets, as discussed in the previous sections. Further, an estimate can be made based on TAS, which is a popular tool to study charge generation in OSCs.

\section{Conclusion}
A differential equation model has been introduced which describes the charge generation dynamics in OSCs in order to address the lack of deterministic charge generation models that incorporate all the relevant factors such as morphology, exciton diffusion and interfacial energetics. The model combines exciton diffusion to the donor-acceptor interface with a rate equation model that includes LE, CT states and separated charges. \\
The charge generation efficiencies obtained from the steady-state formulation of the model align with the observed  decrease in internal quantum efficiency in NFA:polymer blends when the driving force for exciton dissociation approaches zero \cite{Zhou.2021}\cite{Pranav.2024}, corresponding to equal LE and CT energies. Although charge generation efficiencies derived from standard rate models reflect the general efficiency trend for reasonable domain sizes when a reduction factor in the exciton dissociation rate is incorporated, these models do not precisely match the efficiencies obtained from our model. This emphasises that a combined exciton diffusion-dissociation model should be used as an alternative to the established rate models in future OSC research, e.g.\ for the analysis of charge generation efficiencies but also for photo- or electroluminescence. Moreover, our analysis suggests that a substantial diffusion length based on a prolonged intrinsic lifetime enables near unity charge generation efficiencies despite a moderate domain size and a diminishing or even negative energetic offset, the latter being essential to reduce $V_\mathrm{OC}$ losses. This conclusion is corroborated by both experimental findings \cite{Classen.2020} and recent theoretical work \cite{Riley.2022}, thereby validating our modelling approach. \\
Our model is capable of simulating time-resolved experiments, such as TAS or transient photoluminescence, and can be used to disentangle the experimental outcomes with regard to the underlying processes. Our key finding is that the reported time to complete HT in low-offset NFA-based OSCs ($\sim100\,$ps) can be considerably longer than the pure diffusion time to the donor-acceptor interface. A finite exciton dissociation rate at the interface significantly slows down charge generation. Therefore, domain size estimates based on the assumption that excitons are instantaneously quenched at the donor-acceptor interface are only valid for high driving forces $\Delta E_\mathrm{LE-CT}  > 0.2\,$eV, which means that the CT state lies well below the LE state. Furthermore, the model is a good fit to experimental TAS data of PM6:Y6 and provides an estimate of the domain size of Y6 that is in agreement with previously reported values in the literature. \\
In summary, our model can be employed to comprehend and examine steady-state as well as dynamic processes in OSCs. For systems where the driving force for charge transfer is reduced and the assumption of immediate exciton quenching at the donor-acceptor interface is therefore not met, the model can also be employed as an alternative method for estimating domain sizes. We believe that our work will contribute to a deeper understanding of the charge generation dynamics in OSCs, thereby facilitating further improvements in their efficiency.

\section*{Materials and Methods}
\begin{itemize}
	\item Materials and sample preparation: PM6 and Y6 were both obtained from 1-Materials Inc. The full chemical names are given in the supplementary material (\textbf{\ref{sup_materials}}). For the BHJ samples, blend solutions were prepared with a total concentration of $14\,\frac{\mathrm{mg}}{\mathrm{ml}}$ using chloroform ($\mathrm{CHCL}_3$) as the solvent (from Carl Roth), and a weight ratio of $1$:$1.2$. The solution was stirred for $3$ hours in a nitrogen-filled glovebox. The blend was then spin-coated onto glass substrates at an optimized speed of $1000$ RPM, resulting in a photoactive layer with a thickness of $100-110\,$nm.
	\item Transient absorption spectroscopy: Femtosecond transient spectroscopy measurements were conducted using a custom-built setup. The output from an amplified Ti:Sapphire laser (Libra, Coherent, $800\,$nm, $1\,$KHz) was coupled with an Optical Parametric Amplifier (Opera Solo, Coherent) to produce pump laser pulses ($50\,$fs, $720\,$nm), which were chopped at $\frac{\omega}{2}$ ($500\,$Hz). A portion of the fundamental laser beam was focused on an undoped YAG crystal to generate a broadband white light continuum. The probe pulses were spectrally dispersed by prism spectrometers and collected using either a CMOS camera (for the visible spectrum) or InGaAs photodiode arrays (for the IR spectrum). To reduce pulse-to-pulse fluctuations, the probe continuum beam was split using a broadband beam splitter before the sample, allowing shot-to-shot monitoring of probe intensity. The pump-probe polarization was set at the magic angle ($54.7\,^\circ$) to avoid orientation-dependent effects. Differential absorption signals were determined by comparing sequential probe shots with the pump on and off. Time delays of up to $\sim 7\,$ns were achieved using a retroreflector mounted on a computer-controlled translational stage. For each time delay, $6000$ shots were averaged, with each measurement repeated for at least three scans. The resulting data, saved as .dat files, were processed using MATLAB for background subtraction and chirp correction. All measurements were performed under a nitrogen atmosphere to prevent degradation due to air exposure.
\end{itemize}

\section*{Acknowledgements}
The authors acknowledge the funding support from the Deutsche Forschungsgemeinschaft (DFG, German Research Foundation) through the project Extraordinaire (Project Number 460766640).

\printbibliography

\makeatletter\@input{aux_alternative_supplementary.tex}\makeatother
\end{document}


\maketitle

\section{Supplementary Notes}
\subsection{The Exciton Dissociation Constant $\kappa_\mathrm{diss,LE}$} \label{sup_kappa_dissex}
In the main text, the CT state density is defined in terms of a surface density. An alternative approach would be to assume that the CT states are localised in a layer with thickness $d$. In the case of our spherically symmetric model, which we will employ from this point onward in this section, this would entail that the CT states are described by a volume density, $n_\mathrm{CT,V}(r,t)$, with $R \leq r \leq R+d$. The total volume available for CT excitons is then given by $V_\mathrm{CT} \approx 4 \pi R^2 d$. We will assume that the CT excitons are uniformly distributed in this volume, so that $n_\mathrm{CT,V} = n_\mathrm{CT,V}(t)$. We will again make use of a Robin-type boundary condition to describe charge generation at the interface, which changes accordingly to
\begin{align}
	-D \frac{\partial n_\mathrm{LE}}{\partial r}(R,t) = \kappa_\mathrm{diss,LE} n_\mathrm{LE}(R,t) - \kappa_\mathrm{ref} n_\mathrm{CT,V}(t).
\end{align}
Note that both $\kappa_\mathrm{diss,LE}$ and  $\kappa_\mathrm{ref}$ must have units of a velocity. We will leave these constants unspecified for now. Since the units of the CT density have changed, the corresponding rate equation must be adjusted:
\begin{align} \label{sup_CT_volume}
	\frac{d n_\mathrm{CT,V}}{d t}(t) = k_\mathrm{diss,LE} n_\mathrm{LE}(R,t) - (k_\mathrm{ref} + k_\mathrm{f,CT}+k_\mathrm{diss,CT})n_\mathrm{CT,V}(t)
\end{align}
where $k_\mathrm{diss,LE/ref}$ are \enquote{real} rates with unit one over time. The constants $\kappa_\mathrm{diss,LE/ref}$ can be determined by ensuring the conservation of excitons as demonstrated in equation \eqref{modelling_conservation} of the main text. This leads to 
\begin{align}
	n_\mathrm{LE}(R,t)(V_\mathrm{CT} k_\mathrm{diss,LE}-4 \pi R^2 \kappa_\mathrm{diss,LE}) + n_\mathrm{CT,V}(t)(4 \pi R^2 \kappa_\mathrm{ref} - V_\mathrm{CT} k_\mathrm{ref}) \stackrel{!}{=} 0 
\end{align}
implying that $\kappa_\mathrm{diss,LE/ref} \stackrel{!}{=} k_\mathrm{diss,LE/ref}d$. Multiplying \eqref{sup_CT_volume} with $d$ yields
\begin{align}
	-D \frac{\partial n_\mathrm{LE}}{\partial r}(R,t) &= \kappa_\mathrm{diss,LE} n_\mathrm{LE}(R,t) - k_\mathrm{ref} d n_\mathrm{CT,V}(t) \\
	\frac{d (n_\mathrm{CT,V}d)}{d t}(t) &= \kappa_\mathrm{diss,LE} n_\mathrm{LE}(R,t) - (k_\mathrm{ref} + k_\mathrm{f,CT}+k_\mathrm{diss,CT})n_\mathrm{CT,V}(t)d.
\end{align}
By defining $n_\mathrm{CT}(t) = n_\mathrm{CT,V}(t)d$ as a surface density we will return to our original model where the thickness of the CT layer is included in the exciton dissociation constant $\kappa_\mathrm{diss,LE} = k_\mathrm{diss,LE}d$. \\
This interpretation of $d$ is also applicable when considering the system in thermal equilibrium. We are now returning to the model presented in the main text, which includes a CT surface density. In thermal equilibrium, the densities of LE and CT can be estimated by Boltzmann statistics 
\begin{align}
	n_\mathrm{LE} \propto N_\mathrm{LE} \exp{\left( -\frac{E_\mathrm{LE}}{k_\mathrm{B}T} \right)}, \quad n_\mathrm{CT} \propto N_\mathrm{CT} \exp{\left( -\frac{E_\mathrm{CT}}{k_\mathrm{B}T} \right)}.
\end{align}
where $N_\mathrm{LE}$ represents the density of states of singlet excitons per volume, whereas $N_\mathrm{CT}$ denotes the density of states of CT excitons per area. In thermal equilibrium the total flux must vanish, implying that also the flux at the interface \eqref{modelling_diss_boundary} must be zero. When using $k_\mathrm{diss,LE/ref}$ as outlined by Marcus theory \eqref{steady_Marcus}, the zero flux condition at the interface yields 
\begin{align}
	N_\mathrm{LE}d = N_\mathrm{CT}
\end{align}
which makes sense when viewing $d$ as the thickness of the CT layer and assuming equal volume densities of states for LE and CT.
\subsection{Numerical Solution of the Coupled Partial Differential Equation - Ordinary Differential Equation System} \label{sup_numerics}
We use a simple method of lines (MOL) approach together with central finite differences to approximate the system defined by equations \eqref{modelling_sphere_diffu} to \eqref{modelling_sphere_IC} with a pure ordinary differential equation (ODE) system. We define $N$ points with equal spacing  
\begin{align}
	r_0 = 0 < r_1 < \dots < r_{N-2} < r_{N-1} = R
\end{align}
so that the difference $\Delta r \coloneq r_{i+1}-r_i$ is constant. The exact solution at the specified points $n_\mathrm{LE}(r_i,t)$ is approximated by $n_i(t)$. The Laplace operator is discretised by central finite differences (for further details, see \cite{Langtangen.2017}\cite{Schiesser.2009}\cite{Versypt.2014}). Care must be taken at the origin and at the boundary. Applying L'Hosptial's rule yields the radial component of the Laplace operator at the origin $\Delta = 3\frac{\partial^2}{\partial r^2}$. The missing fictitious value $n_{-1}$ required to approximate derivatives at the origin can be extracted from the first boundary condition in \eqref{modelling_sphere_boundary}, $n_{-1}=n_1$. Similarly, the second boundary condition is used to extract the fictitious $n_N$
\begin{align}
	n_N = -\frac{2 \Delta r}{D} \kappa_\mathrm{diss,LE} n_{N-1} + \frac{2 \Delta r}{D} k_\mathrm{ref} n_\mathrm{CT} + n_\mathrm{N-2}.
\end{align}
The initial conditions for the discretised partial differential equation are $n_i(0)=n_\mathrm{LE,0}(r_i)$. Altogether the ODE system reads
\begin{align}
	&\frac{dn_i}{dt} = \begin{cases}
		\frac{6D}{\Delta r^2}(n_1-n_0) - R_\mathrm{LE}(n_0) + G_\mathrm{LE}, & \mathrm{if}~i = 0 \\
		\frac{2D}{\Delta r^2}(n_{N-2}-n_{N-1}) + 2\left(\frac{1}{R}+\frac{1}{\Delta r} \right)(k_\mathrm{ref}n_\mathrm{CT} - \kappa_\mathrm{diss,LE} n_{N-1}) - R_\mathrm{LE}(n_{N-1}) + G_\mathrm{LE}, & \mathrm{if}~i = N-1 \\
		\frac{D}{i \Delta r^2}(n_{i+1}(i+1)-2n_i i + n_{i-1}(i-1)) - R_\mathrm{LE}(n_i) + G_\mathrm{LE}, & \mathrm{otherwise}
	\end{cases} \label{sup_PDE_MOL} \\
	&\frac{dn_\mathrm{CT}}{dt} = \kappa_\mathrm{diss,LE}n_{N-1} + (k_\mathrm{ref}+k_\mathrm{f,CT}+k_\mathrm{diss,CT})n_\mathrm{CT} \\
	& \frac{dN_\mathrm{CT}}{dt} = k_\mathrm{diss,CT} 4\pi R^2 n_\mathrm{CT} \\
	&n_i(0) = n_{\mathrm{LE},0}(r_i), \quad n_{\mathrm{CT}}(0) = 0, \quad N_\mathrm{CS}(0) = 0.
\end{align} 
Note that that the second equation in \eqref{sup_PDE_MOL} implements a reflecting boundary when $k_\mathrm{ref}$ and $\kappa_\mathrm{diss,LE}$ are set to zero. We are using the implicit Runge-Kutta method \enquote{Radau} \cite{Hairer.1991}\cite{Virtanen.2020} to solve this system. 
\subsection{Planar Geometry} \label{sup_planar}
The equations that describe the planar geometry presented in the main text are
\begin{align} \label{modelling_planar}
	&\frac{\partial n_{\mathrm{LE}}}{\partial t}(x,t) = \frac{\partial^2n_{\mathrm{LE}}}{\partial x^2}(x,t) + G_\mathrm{LE}(x,t) - R_\mathrm{LE}(x,t), \quad 0 \leq x \leq L \\ 
	&\frac{\partial n_{\mathrm{LE}}}{\partial x}(0,t) = 0, \quad -D \frac{\partial n_\mathrm{LE}}{\partial x}(L,t) = \kappa_\mathrm{diss,LE} n_\mathrm{LE}(L,t) - k_\mathrm{ref} n_\mathrm{CT}(t) \\
	&\frac{d n_\mathrm{CT}}{d t}(t) = \kappa_\mathrm{diss,LE} n_\mathrm{LE}(L,t) - (k_\mathrm{ref} + k_\mathrm{f,CT}+k_\mathrm{diss,CT})n_\mathrm{CT}(t)  \\
	&\frac{dN_\mathrm{CS}}{dt}(t) = k_\mathrm{diss,CT}A n_\mathrm{CT}(t) \\
	&n_\mathrm{LE}(x,0) = n_\mathrm{LE,0}(x), \quad n_\mathrm{CT}(0) = n_{\mathrm{CT},0}, \quad N_\mathrm{CS}(0) = N_{\mathrm{CS},0}. 
\end{align} 
Equation \eqref{modelling_sphere_CT_sol} is applicable for the planar system as well when $R$ is replaced by $L$. Therefore, a numerical solution is required. The same expressions for $n_{-1}$ and $n_N$ are used and the Laplace operator is approximated by the expression $\frac{D}{\Delta x^2}(n_{i+1}-2n_i+n_{i+1})$.
\subsection{Decay Efficiencies} \label{sup_decay_eff}
Similar to the charge generation efficiency $\eta_\mathrm{gen,CS}$, the efficiencies of singlet, $\eta_\mathrm{f,LE}$, and CT decay, $\eta_\mathrm{f,CT}$, can be determined
\begin{align}
	\eta_\mathrm{f,CT} &= \frac{k_\mathrm{f,CT}4 \pi R^2 n_\mathrm{CT}}{\frac{4}{3}\pi R^3 G} = \frac{3 k_\mathrm{f,CT}n_\mathrm{CT}}{RG} \notag \\
	&= 3\frac{\kappa_\mathrm{diss,LE}}{R k_\mathrm{f,LE}} P_\mathrm{f,CT} \frac{1}{1+\frac{\frac{\kappa_\mathrm{diss,LE}}{R k_\mathrm{f,LE}}(1-P_\mathrm{ref})\frac{R}{L_\mathrm{D}}\tanh{(\frac{R}{L_\mathrm{D}})}}{1-\frac{L_\mathrm{D}}{R}\tanh{(\frac{R}{L_\mathrm{D}})}}} \\
	P_\mathrm{f,CT} &\coloneq \frac{k_\mathrm{f,CT}}{k_\mathrm{ref}+k_\mathrm{f,CT}+k_\mathrm{diss,CT}} \\
	\eta_\mathrm{f,LE} &= \frac{k_\mathrm{f,LE}4 \pi \int_0^Rr^2n_\mathrm{LE}(r)dr }{\frac{4}{3}\pi R^3 G} = \frac{3 k_\mathrm{f,LE}\int_0^Rr^2n_\mathrm{LE}(r)dr}{R^3G} \notag \\ 
	&= 1 - 3\frac{\kappa_\mathrm{diss,LE}}{R k_\mathrm{f,LE}} (1-P_\mathrm{ref}) \frac{1}{1+\frac{\frac{\kappa_\mathrm{diss,LE}}{R k_\mathrm{f,LE}}(1-P_\mathrm{ref})\frac{R}{L_\mathrm{D}}\tanh{(\frac{R}{L_\mathrm{D}})}}{1-\frac{L_\mathrm{D}}{R}\tanh{(\frac{R}{L_\mathrm{D}})}}}. \label{sup_LE_decay_eff}
\end{align}
The here defined $P_\mathrm{f,CT}$ is the probability for CT excitons decaying to the ground state. Since $P_\mathrm{f,CT}+P_\mathrm{diss,CT} = 1-P_\mathrm{ref}$ we have $\eta_\mathrm{gen,CS} + \eta_\mathrm{f,CT} + \eta_\mathrm{f,LE} = 1$. So every generated exciton either decays via the LE or CT states or dissociates into free charges, confirming the result of \eqref{modelling_conservation} in the main text.
\subsection{Charge Generation Efficiency - Reference Model} \label{sup_ref_model}
We compare the calculated charge generation efficiency from our spherical model with a simple reference model (e.g.\ used in \cite{Pranav.2024})
\begin{align} 
	\frac{d[n_\mathrm{LE}]}{dt} &= G - (k_\mathrm{diss,LE} + k_\mathrm{f,LE} )[n_\mathrm{LE}] + k_\mathrm{ref} [n_\mathrm{CT}] \label{sup_ref_model_LE}\\
	\frac{d[n_\mathrm{CT}]}{dt} &= k_\mathrm{diss,LE}[n_\mathrm{LE}] - (k_\mathrm{ref} + k_\mathrm{f,CT} + k_\mathrm{diss,CT})[n_\mathrm{CT}] \label{sup_ref_model_CT}
\end{align}
where $[n_\mathrm{LE}]$ and $[n_\mathrm{CT}]$ refer to averaged volume densities. The process of diffusion is typically accounted for implicitly by weighting the exciton dissociation rate with a prefactor of $0.1$, $k_\mathrm{diss,LE} \mapsto 0.1 \cdot k_\mathrm{diss,LE}$ (see the main text for further details on the physical justification). The charge generation efficiency for this model can be obtained by calculating the steady-state solutions $[n_\mathrm{LE,equ}]$ and $[n_\mathrm{CT,equ}]$ of the system \eqref{sup_ref_model_LE}, \eqref{sup_ref_model_CT}. The charge generation efficiency $\eta_\mathrm{gen,CS}$ is then 
\begin{align}
	\eta_\mathrm{gen,CS} &= \frac{k_\mathrm{diss,CT}[n_\mathrm{CT,equ}]}{G} \notag \\
	& = \frac{k_\mathrm{diss,CT}k_\mathrm{diss,LE}}{k_\mathrm{diss,LE}(k_\mathrm{diss,CT}+k_\mathrm{f,CT})+k_\mathrm{f,LE}(k_\mathrm{ref}+k_\mathrm{diss,CT}+k_\mathrm{f,CT})}.
\end{align}
\subsection{The Mean Survival Time of Excitons} \label{sup_mean_times}
\begin{figure}[H]
	\centering
	\includegraphics[width = 1 \textwidth]{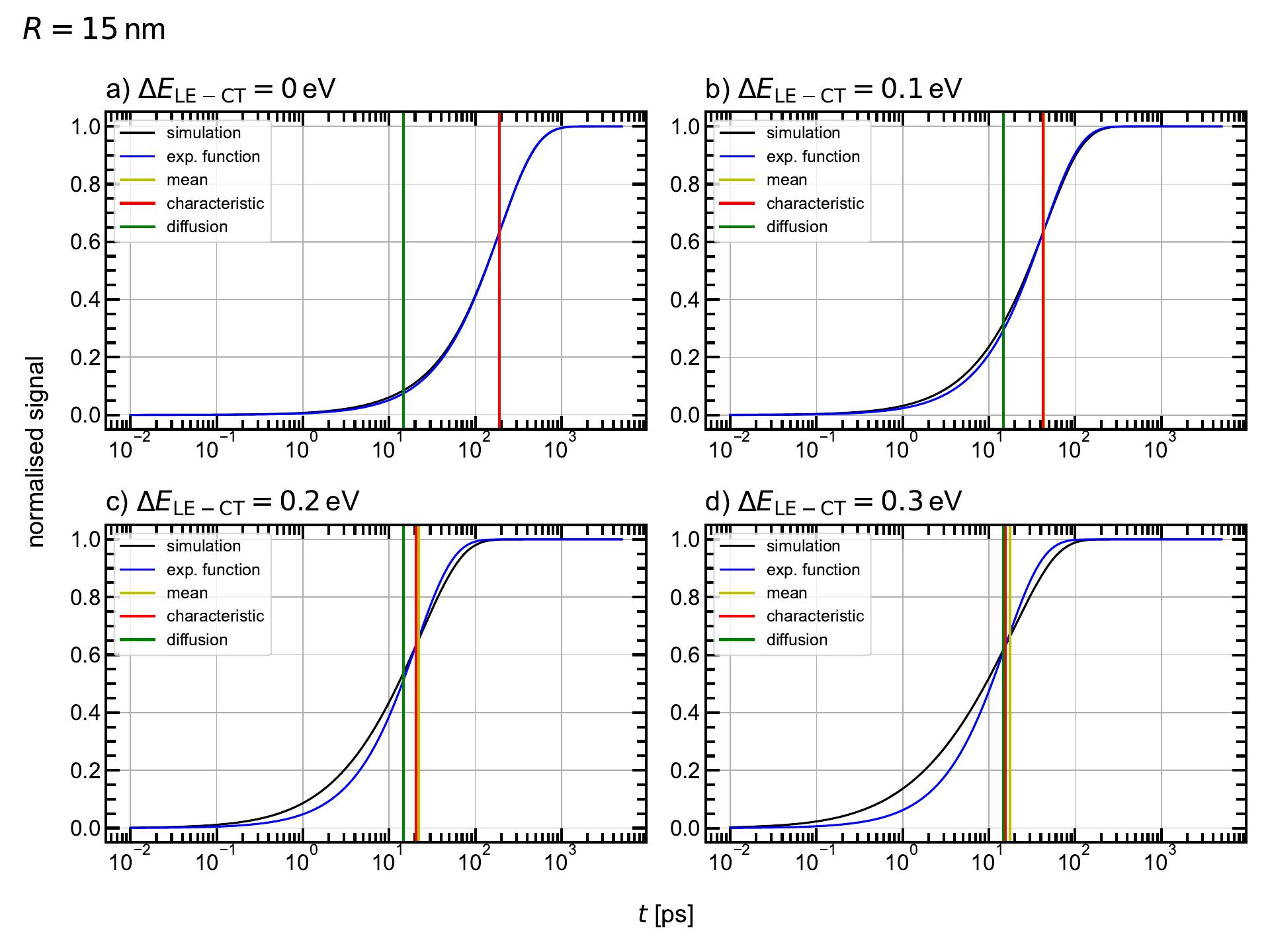}
	\caption{The simulated GSB is normalised to its maximum value and compared to an exponential function for selected offsets and a domain radius of $R=15\,$nm. The same parameters are used as in table \textbf{\ref{tab:selected_times}}. The exponential function $g$ is constructed according to $g(t) = 1 - \exp\left(\frac{t}{\tau_\mathrm{GSB}}\right)$, where $\tau_\mathrm{GSB}$ represents the characteristic GSB rise time defined in the main text. By definition, $g$ and the normalised GSB signal intersect at a value of $1-e^{-1} \approx 63\,\%$. The characteristic time and the mean time of charge generation, as defined in \textbf{\ref{sup_mean_times}}, are illustrated by the coloured vertical lines (the numerical values are presented in table \textbf{\ref{tab:selected_times}}). For small offsets, the constructed exponential closely approximates the GSB kinetics. Consequently, the GSB rise time and the mean charge generation time coincide. As the offset increases, the exponential begins to deviate from the GSB. This is also evident in the difference between the characteristic GSB time and the mean charge generation time. The survival time from table \textbf{\ref{tab:results_MFPT}} 3) is also shown and labelled as diffusion time. The characteristic GSB and mean charge generation times approach this diffusion time when going to higher driving forces for exciton dissociation.}
	\label{fig:sup_GSB_vs_exp_15}
\end{figure}

\begin{figure}[H]
	\centering
	\includegraphics[width = 1 \textwidth]{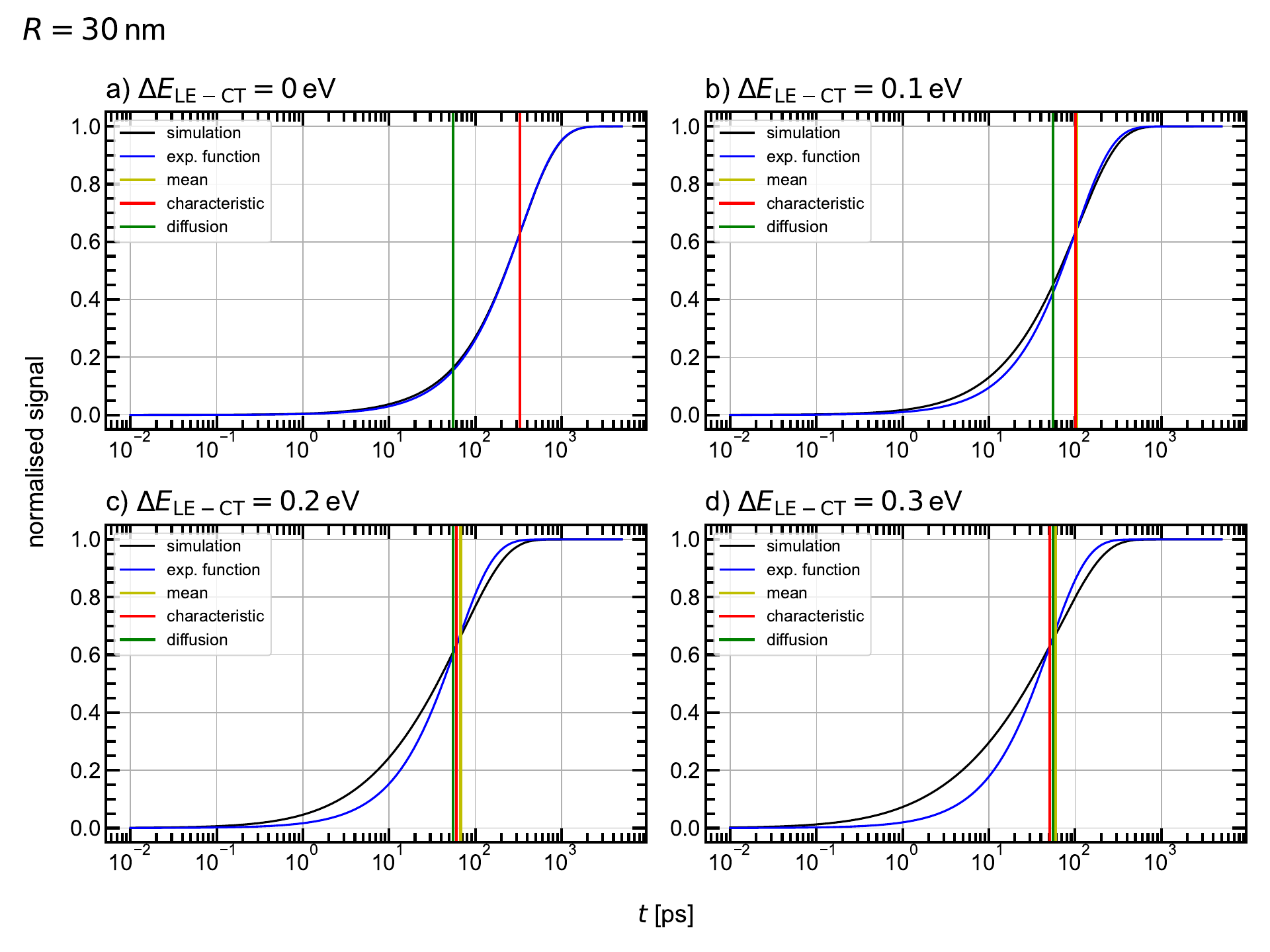}
	\caption{The simulated GSB is normalised to its maximum value and compared to an exponential function for selected offsets and a domain radius of $R=30\,$nm. The same parameters are used as in table \textbf{\ref{tab:selected_times}}. The exponential function $g$ is constructed according to $g(t) = 1 - \exp\left(\frac{t}{\tau_\mathrm{GSB}}\right)$, where $\tau_\mathrm{GSB}$ represents the characteristic GSB rise time defined in the main text. By definition, $g$ and the normalised GSB signal intersect at a value of $1-e^{-1} \approx 63\,\%$. The characteristic time and the mean time of charge generation, as defined in \textbf{\ref{sup_mean_times}}, are illustrated by the coloured vertical lines (the numerical values are presented in table \textbf{\ref{tab:selected_times}}). For small offsets, the constructed exponential closely approximates the GSB kinetics. Consequently, the GSB rise time and the mean charge generation time coincide. As the offset increases, the exponential begins to deviate from the GSB. This is also evident in the difference between the characteristic GSB time and the mean charge generation time. The survival time from table \textbf{\ref{tab:results_MFPT}} 3) is also shown and labelled as diffusion time. The characteristic GSB and mean charge generation times approach this diffusion time when going to higher driving forces for exciton dissociation. The characteristic GSB time becomes even less than the diffusion time, while the mean charge generation time remains above it.}
	\label{fig:sup_GSB_vs_exp_30}
\end{figure}

The formalism presented here is based on the approach summarised in \cite{Wurschum.2023}. We use the probabilistic interpretation of the diffusion equation and assume that one single exciton is generated in a domain $\Omega \subset \mathbb{R}^3$ that is defined in the same manner as in the main text. Let $n$ be the probability density describing the probability of finding the exciton in some volume $dV$ inside the domain. As only one exciton is considered and exciton-exciton interactions are thereby neglected, the time evolution of $n$ is given by 
\begin{align}
	\frac{\partial n}{\partial t} = D \Delta n - k_\mathrm{f,LE} n, \quad \mathrm{on}~\Omega.
\end{align} 
$D$ represents the diffusion coefficient, while $k_\mathrm{f,LE}$ denotes the exciton decay rate. For the purposes of this analysis, we will ignore the possibility of reforming an exciton from the CT state, as our focus is on the initial diffusion process of the exciton to the donor-acceptor interface. The boundary condition reads
\begin{align}
	-D\nabla n \cdot \boldsymbol{\nu} = \kappa_\mathrm{diss,LE} n, \quad \mathrm{on}~\partial \Omega.
\end{align} 
where $\boldsymbol{\nu}$ is again the outward unit normal vector on $\partial \Omega$ and $\kappa_\mathrm{diss,LE}$ denotes the exciton dissociation constant at the interface. $\kappa_\mathrm{diss,LE} = \infty$ describes an interface that facilitates immediate exciton dissociation (equivalent to a Dirichlet boundary condition which constrains $n$ to be zero at the boundary), while $\kappa_\mathrm{diss,LE} = 0$ accounts for a reflective boundary. The initial condition $n_0$ must fulfil 
\begin{align}
	\int_\Omega n_0(\mathbf{x}) dV = 1
\end{align}
because the exciton shall be located in the domain initially with probability one. Similarly, the time dependent survival probability $S$ of the exciton can be evaluated by the following integrals
\begin{align}
	S(t) &= \int_\Omega n(\mathbf{x},t) dV \notag \\
	&= \int_t^\infty f(\tau) d\tau \label{sup_definition_f}
\end{align}
where we define $f(t)dt$ as the probability that the exciton is quenched during the time interval $[t,t+dt]$, either via dissociation following diffusion to the interface or via decay in the bulk. It is evident from \eqref{sup_definition_f} that $f(t) = -\frac{dS}{dt}(t)$, since $f(\infty)$ must be zero. The probability density $f$ facilitates the calculation of the mean exciton survival time
\begin{align} \label{sup_def_mean}
	\langle t \rangle_f &= \int_0^\infty t f(t) dt \notag \\
	&=-\int_0^\infty t \frac{dS}{dt}(t) dt \notag \\
	&= - \left[t S(t)\right]_0^\infty + \int_0^\infty \int_\Omega n(\mathbf{x},t) dV dt \notag \\
	&= \int_\Omega \int_0^\infty  n(\mathbf{x},t) dt dV 
\end{align}
where we have used integration by parts, $S(\infty) = 0$ and Fubini's theorem. The inner integral in the last line of \eqref{sup_def_mean} is the Laplace transform of $n$ evaluated at zero
\begin{align}
	\Tilde{n}(\mathbf{x}) \coloneq \mathcal{L}\{n\}(\mathbf{x},0) = \int_0^\infty  n(\mathbf{x},t) dt
\end{align}
and the differential equation for $\Tilde{n}$ reads
\begin{align}
	D \Delta \Tilde{n} - k_\mathrm{f,LE} \Tilde{n} + n_0 &= 0, \quad \mathrm{on}~\Omega \\
	-D\nabla \Tilde{n} \cdot \boldsymbol{\nu} &= \kappa_\mathrm{diss,LE} \Tilde{n}, \quad \mathrm{on}~\partial \Omega \label{sup_boundary_laplace}.
\end{align}
The resulting equations are essentially identical to those of the steady-state formulation of the model under constant generation when reformation is neglected. The mean survival time can now be obtained through integration $\langle t \rangle_f = \int_\Omega \Tilde{n}(\mathbf{x}) dV$. \\
To estimate the mean survival times of our BHJ model, we adopt the spherical geometry with radius $R$. The boundary condition is then given by $-D\frac{\partial \Tilde{n}}{\partial r}(R) = \kappa_\mathrm{diss,LE} \Tilde{n}(R)$. We again assume that the probability of generating an exciton is uniform across the domain, which leads to the initial condition $n_0 = \frac{3}{4 \pi R^3}$. Consequently, the system is spherically symmetric, and therefore only the radial component needs to be considered. 
\begin{table}[h!] 
	\centering
	\caption{The mean time of leaving the system $\langle t \rangle_f$ for different combinations of $k_\mathrm{f,LE}$ and $\kappa_\mathrm{diss,LE}$. $L_\mathrm{D} \coloneq \sqrt{\frac{D}{k_\mathrm{f,LE}}}$ is the diffusion length.}
	\label{tab:results_MFPT}
	\begin{tabular}{cc|c}
		& & $\langle t \rangle_f = $ \\
		\hline
		1) & $k_\mathrm{f,LE} = 0$, $\kappa_\mathrm{diss,LE} = \infty$ & $\frac{2}{5} \cdot \frac{R^2}{6D}$ \\
		2) & $k_\mathrm{f,LE} = 0$, $\kappa_\mathrm{diss,LE} \neq \infty$ & $\frac{2}{5} \cdot \frac{R^2}{6D} + \frac{R}{3 \kappa_\mathrm{diss,LE}}$ \\
		3) & $k_\mathrm{f} \neq 0$, $\kappa_\mathrm{diss,LE} = \infty$ & $\frac{1}{k_\mathrm{f,LE}} \left[1 - 3 \left( \frac{L_\mathrm{D}}{R} \right)^2 \left( \frac{R}{L_\mathrm{D}} \coth\left(\frac{R}{L_\mathrm{D}} \right) -1 \right) \right]$ \\
		4) & $k_\mathrm{f,LE} \neq 0$, $\kappa_\mathrm{diss,LE} \neq \infty$ & $\frac{1}{k_\mathrm{f,LE}}\left[1 - \frac{3 \kappa_\mathrm{diss,LE}}{k_\mathrm{f,LE} R} \frac{1}{1+\frac{\tanh\left(\frac{R}{L_\mathrm{D}}\right)\frac{\kappa_\mathrm{diss,LE}}{k_\mathrm{f,LE}R}\frac{R}{L_\mathrm{D}}}{1-\frac{L_\mathrm{D}}{R}\tanh\left(\frac{R}{L_\mathrm{D}}\right)}} \right]$	  	 
	\end{tabular}
\end{table}
The results for our selected geometry and different combinations of $k_\mathrm{f,LE}$ and $\kappa_\mathrm{diss,LE}$ are summarised in table \textbf{\ref{tab:results_MFPT}}. As before, we have defined the diffusion length as follows without the prefactor of $\sqrt{6}$, $L_\mathrm{D} \coloneq \sqrt{\frac{D}{k_\mathrm{f,LE}}}$. In the case where the decay rate $k_\mathrm{f,LE}$ is set to zero, excitons can only be quenched by dissociation following diffusion to the interface. When the exciton dissociation constant $\kappa_\mathrm{diss,LE}$ is set to  infinity, all trajectories reaching the interface end immediately. Both special cases together yield the mean first passage time (MFPT) of the exciton with respect to the interface, which is presented in the first line of table \textbf{\ref{tab:results_MFPT}} and in the main text \eqref{times_MFPT}. When the exciton dissociation time is finite, the mean diffusion time increases by $\frac{R}{3 \kappa_\mathrm{diss,LE}}$. This is the reason why the observed rise times of the GSB are considerably longer than the pure diffusion time. The additional component scales with the radius $R$, which underlines that the discussed interplay between diffusion and dissociation is more important for large domains. The third line of the table presents the expression when the MFPT is reduced by the finite intrinsic exciton lifetime. This is analogous to the scenario in which charge generation is solely limited by the decay of excitons in the bulk. We use this time as the lower bound of the characteristic GSB time in figures \textbf{\ref{fig:times}b)} and \textbf{\ref{fig:times}c)} in the main text. The final line of the table displays the general case. All the other expressions can be obtained as limiting cases from this one. It should be noted that the mean survival time 4) in table \textbf{\ref{tab:results_MFPT}} is essentially the intrinsic lifetime $\tau$ weighted by the efficiency of singlet decay \eqref{sup_LE_decay_eff} when the reformation rate is set to zero. \\
An additional potential methodology for estimating the average domain size in a blend is to monitor the exciton population following pulsed excitation using a time-resolved method, such as transient photoluminescence. Subsequently, the mean exciton survival time can be calculated from the experimental data. When additionally the intrinsic lifetime $k_\mathrm{f,LE}^{-1}$, the diffusion coefficient $D$ and $\kappa_\mathrm{diss,LE}$ (via the offset $\Delta E_\mathrm{LE-CT}$ and Marucs theory) are known, the radius $R$ is the only remaining  unknown quantity. The final or the second (suitable if $k_\mathrm{f,LE}^{-1} \gg \langle t \rangle_f$) equation of table \textbf{\ref{tab:results_MFPT}} can then be used to estimate $R$. However, the experiment must be conducted at low fluences, since SSA is not included in the above expression. Furthermore, CT dissociation should be efficient to suppress the effect of reformation in low-offset systems. \\
In figure \textbf{\ref{fig:times}c)} of the main text, the characteristic GSB rise time falls below the diffusion time for high offsets $\Delta E_\mathrm{LE-CT}$. We have attributed this to the fact that the GSB rise does not follow an exponential function in a strict manner, as illustrated by figures \textbf{\ref{fig:sup_GSB_vs_exp_15}} and \textbf{\ref{fig:sup_GSB_vs_exp_30}}. In the following we will demonstrate that the mean charge generation time is in fact bounded from below by equation 3) in table \textbf{\ref{tab:results_MFPT}}. To this end, the GSB signal is normalised to its maximum value like in the figures below (\textbf{\ref{fig:sup_GSB_vs_exp_15}} and \textbf{\ref{fig:sup_GSB_vs_exp_30}}) and interpreted as a cumulative probability distribution function. Consequently, the derivative of our normalised GSB defines a probability density $f$. It should be acknowledged that the function $f$ is not well-defined in this case, given that the GSB population can decrease in our model, which leads to negative derivatives. We perform the analysis anyway, because the GSB kinetics are almost non-decreasing with the parameters present in figure \textbf{\ref{fig:times}c)} (in fact, the GSB drops only non-noticeably for $t>1\,$ns). The mean time is calculated as in the first line of equation \eqref{sup_def_mean}. Numeric values of the characteristic GSB time (see main text), the mean time defined above and the survival times of \textbf{\ref{tab:results_MFPT}} are provided in table \textbf{\ref{tab:selected_times}} for selected offsets.
\begin{table}[h!] 
	\centering
	\caption{The characteristic GSB time defined in the main text and the mean time of charge generation for different combinations of $R$ and $\Delta E_\mathrm{LE-CT}$. The MFPT is calculated by \textbf{\ref{tab:results_MFPT}} 1) and the reduced MFPT by \textbf{\ref{tab:results_MFPT}} 3). The following parameters are used: $k_\mathrm{f,ex}=k_\mathrm{f,CT}=\frac{1}{1000}\,\mathrm{ps}^{-1}$, $k_\mathrm{diss,CT} = 0.1\,\mathrm{ps}^{-1}$, $D=10^{-2}\,\mathrm{cm}^2\mathrm{s}^{-1}$, $d=1\,$nm, $\lambda=0.45\,$eV, $|H_\mathrm{DA}| = 0.01\,$eV and $T=293.15\,$K.}
	\label{tab:selected_times}
	\begin{tabular}{c|c|c|c|c|c}
		$R$ [nm]& $\Delta E_\mathrm{LE-CT}$ [eV] & char. GSB time [ps] & $ \langle t \rangle_f$ [ps] & MFPT [ps] & reduced MFPT [ps] \\
		\hline
		\multirow{4}{*}{15} & 0 &189.39 & 189.43 & \multirow{4}{*}{15} & \multirow{4}{*}{14.69} \\
		& 0.1 & 42.35 & 42.99 & &  \\
		& 0.2 & 20.43 & 22.05 & &  \\
		& 0.3 & 15.56 & 17.64 & &  \\
		\hline
		\multirow{4}{*}{30} & 0 & 330.30 & 330.69 & \multirow{4}{*}{60} & \multirow{4}{*}{55.28} \\
		& 0.1 & 100.59 & 104.94 & &  \\
		& 0.2 & 60.12 & 67.66 & &  \\
		& 0.3 & 50.99 & 59.61 & & 
	\end{tabular}
\end{table}
The numerical values substantiate that the defined GSB rise time underestimates the mean charge generation time for high energetic offsets, which explains why the GSB rise time is smaller than the defined diffusion time in figure \textbf{\ref{fig:times}}. Conversely, $\langle t \rangle_f$ is well bounded from below by the reduced MFPT. We use the reduced MFPT as a lower bound for the charge generation time, since the finite exciton lifetime precludes charge generation at large timescales, which is not included in the standard MFPT. This is also demonstrated by the numerical values. For $R = 30\,$nm and $\Delta E_\mathrm{LE-CT} = 0.3\,$eV, the mean charge generation time is less than the MFPT but still larger than the MFPT reduced by the finite exciton lifetime. The times of table \textbf{\ref{tab:selected_times}} are visualised in figures \textbf{\ref{fig:sup_GSB_vs_exp_15}} and \textbf{\ref{fig:sup_GSB_vs_exp_30}} by the respective vertical lines.
\subsection{The Charge Generation Efficiency for a Uniform Initial Exciton Distribution} \label{sup_charge_gen_pulsed}
We simulate an experiment where the OSCs is excited by an ultrashort laser pulse at a low fluence.  Hence, exciton-exciton interactions such as SSA can be neglected. This situation is described by the following equations
\begin{align} \label{sup_pulsed exitation}
	&\frac{\partial n_\mathrm{LE}}{\partial t} = D \Delta n_\mathrm{LE} - k_\mathrm{f,LE}n_\mathrm{LE} \quad \mathrm{on}~\Omega \\
	&-D \nabla n_\mathrm{LE} \cdot \boldsymbol{\nu} = \kappa_\mathrm{diss,LE} n_\mathrm{LE} - k_\mathrm{ref} n_\mathrm{CT} \quad \mathrm{on}~\Gamma_\mathrm{d}, \quad \nabla n_\mathrm{LE} \cdot \boldsymbol{\nu} = 0 \quad \mathrm{on}~\Gamma_\mathrm{r} \\
	&\frac{\partial n_\mathrm{CT}}{\partial t} = \kappa_\mathrm{diss,LE} n_\mathrm{LE} - (k_\mathrm{ref} + k_\mathrm{diss,CT}+k_\mathrm{f,CT})n_\mathrm{CT} \quad \mathrm{on}~\Gamma_\mathrm{d} \\
	&n_\mathrm{LE}(\mathbf{x},t=0) = n_{\mathrm{LE},0} \quad \mathbf{x} \in \Omega, \quad n_\mathrm{CT}(\mathbf{x},t=0) = 0 \quad \mathbf{x} \in \Gamma_\mathrm{d} 
\end{align}
where $n_{\mathrm{LE},0}$ represents the uniform initial exciton distribution that is generated by the pump laser pulse. Here we have again assumed that the spatial extent of our sample domain $\Omega$ is sufficiently small, so that changes in the light intensity can be neglected. The total number of generated charges can be calculated by 
\begin{align}
	N_\mathrm{CS,total} &= \int_0^{\infty} \int_{\Gamma_\mathrm{d}} k_\mathrm{diss,CT} n_\mathrm{CT}(\mathbf{x},t) dA dt \notag \\
	&= k_\mathrm{diss,CT} \int_{\Gamma_\mathrm{d}} \int_0^{\infty} n_\mathrm{CT}(\mathbf{x},t) dt dA \notag \\
	&= k_\mathrm{diss,CT} \int_{\Gamma_\mathrm{d}} \Tilde{n}_\mathrm{CT}(\mathbf{x}) dA.
\end{align}
As before, we have made use of Fubini's theorem and defined $\Tilde{n}_\mathrm{CT}$ as the Laplace transform of $n_\mathrm{CT}$ evaluated at zero. Applying the time integration on \eqref{sup_pulsed exitation} yields
\begin{align}
	&0 = D \Delta \Tilde{n}_\mathrm{LE} - k_\mathrm{f,LE}\Tilde{n}_\mathrm{LE} + n_\mathrm{LE,0} \quad \mathrm{on}~\Omega \\
	&-D \nabla \Tilde{n}_\mathrm{LE} \cdot \boldsymbol{\nu} = \kappa_\mathrm{diss,LE} \Tilde{n}_\mathrm{LE} - k_\mathrm{ref} \Tilde{n}_\mathrm{CT} \quad \mathrm{on}~\Gamma_\mathrm{d}, \quad \nabla \Tilde{n}_\mathrm{LE} \cdot \boldsymbol{\nu} = 0 \quad \mathrm{on}~\Gamma_\mathrm{r} \\
	&0 = \kappa_\mathrm{diss,LE} \Tilde{n}_\mathrm{LE} - (k_\mathrm{ref} + k_\mathrm{diss,CT}+k_\mathrm{f,CT})\Tilde{n}_\mathrm{CT} \quad \mathrm{on}~\Gamma_\mathrm{d}.
\end{align}
Here we have defined $\Tilde{n}_\mathrm{LE}(\mathbf{x}) \coloneq \int_0^{\infty} n_\mathrm{LE}(\mathbf{x},t) dt$. The resulting system corresponds to the equations describing the steady-state under constant generation, whereby the generation rate is replaced by $n_\mathrm{LE,0}$. The charge generation efficiency can be obtained by dividing the total number of generated free charges by the initial number of excitons in the domain
\begin{align}
	\eta_\mathrm{gen,CS} = \frac{N_\mathrm{CS,total}}{n_\mathrm{LE,0} \int_\Omega dV} = \frac{k_\mathrm{diss,CT} \int_{\Gamma_\mathrm{d}} \Tilde{n}_\mathrm{CT}(\mathbf{x}) dA}{n_\mathrm{LE,0} \int_\Omega dV}.
\end{align} 
In the case of steady-state, the charge generation efficiency can be obtained in the same manner by replacing $n_\mathrm{LE,0}$ with the constant generation rate. This is the reason why the generation efficiency obtained from the pulsed simulation agrees with the efficiency calculated from steady-state under constant illumination, provided that exciton-exciton interactions are not taken into account.
\subsection{Materials}
\label{sup_materials}
\begin{itemize}
	\item PM6: poly[[4,8-bis[5-(2-ethylhexyl)-4-fluoro-2-thienyl]benzo[1,2-b:4,5-b']dithiophene-2,6-diyl]-2,5- thiophenediyl-[5,7-bis(2-ethylhexyl)-4,8-dioxo-4H,8H-benzo[1,2-c:4,5-c']dithiophene-1,3-diyl]- 2,5-thiophenediyl]
	\item Y6: 2,2'-[[12,13-bis(2-ethylhexyl)-12,13-dihydro-3,9-diundecylbisthieno[2'',3'':4',5']thieno- \newline
	[2',3':4,5]pyrrolo[3,2-e:2',3'-g][2,1,3]benzothiadiazole-2,10-diyl]bis[methylidyne(5,6-difluoro3-oxo-1H-indene-2,1(3H)-diylidene)]]bis[propanedinitrile]
\end{itemize}

\section{Supplementary Figures}
\begin{figure}[H]
	\centering
	\includegraphics[width = 1 \textwidth]{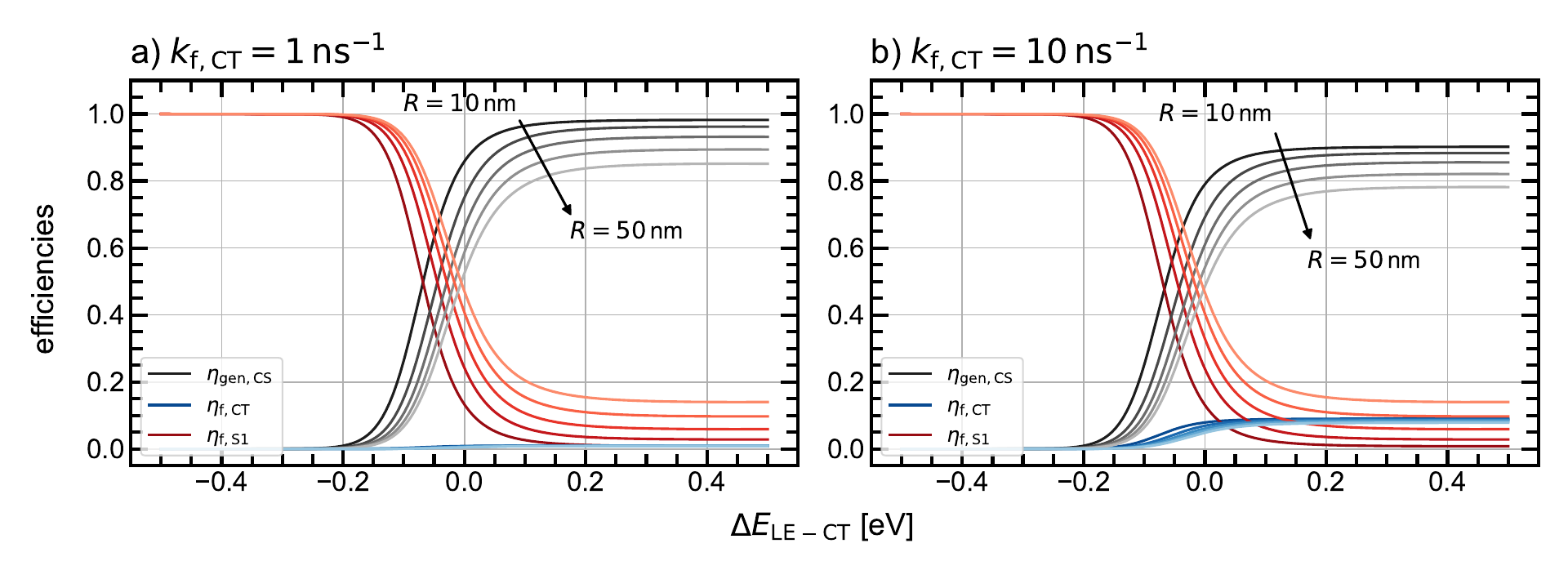}
	\caption{The efficiencies of charge generation $\eta_\mathrm{gen,CS}$, CT decay $\eta_\mathrm{f,CT}$ and singlet exciton decay $\eta_\mathrm{f,LE}$ as a function of the energetic offset using Marcus theory. The domain size is additionally varied from $10\,$nm (dark) to $50\,$nm (light) in steps of $10\,$nm. Further, the following parameters are used: $k_\mathrm{f,LE}=\frac{1}{1000}\,\mathrm{ps}^{-1}$,$D=10^{-2}\,\frac{\mathrm{cm}^2}{\mathrm{s}}$, $k_\mathrm{diss,CT} = 0.1\,\mathrm{ps}^{-1}$, $d=1\,$nm, $\lambda=0.45\,$eV, $|H_\mathrm{DA}| = 0.01\,$eV and $T=293.15\,$K. \\
	\textbf{a)} In the case of high offsets, where the CT energy is significantly lower than that of LE, the majority of generated excitons are converted to free charges. For small positive or even negative $\Delta E_\mathrm{LE-CT}$, exciton dissociation becomes slower while exciton reformation increases. The CT state is then populated only sparsely, which results in a decrease in charge generation efficiency, while almost all excitons decay in the bulk. At vanishing driving force, this loss channel does not shift to CT decay, even when the CT rate is increased as in \textbf{b)}. This is because only a small number of CT states are formed in the first place.}
	\label{fig:sup_efficiencies}
\end{figure}

\begin{figure}[H]
	\centering
	\includegraphics[width = 1 \textwidth]{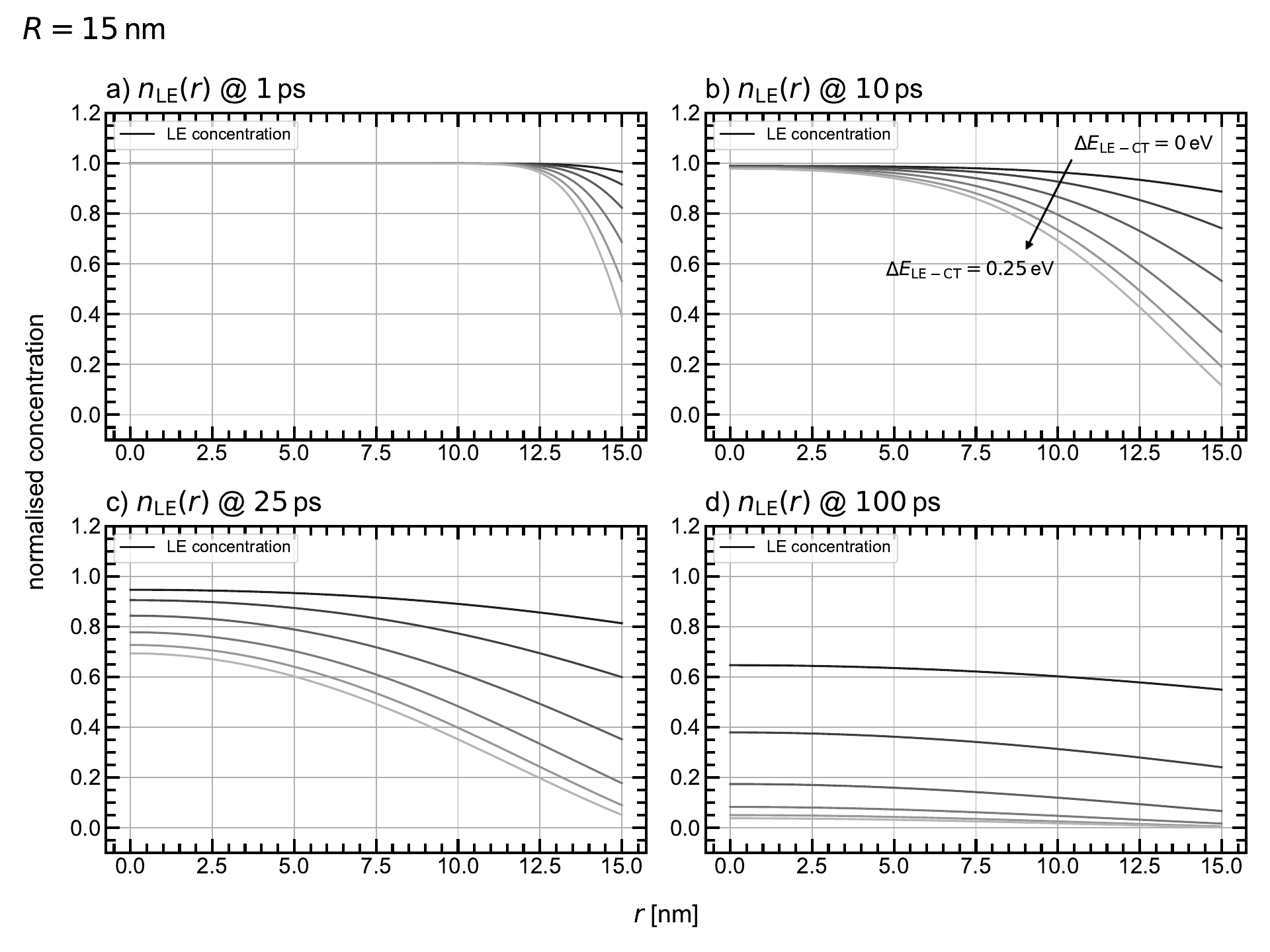}
	\caption{Normalised singlet exciton concentration in a spherical domain with radius $15\,$nm for an initially uniform distribution. The concentration profile is shown for selected delay times after exciton generation. The energetic offset is increased from $0\,$eV (dark) to $0.25\,$eV (light) in steps of $0.05\,$eV, thereby increasing the driving force for exciton dissociation. We consider the low fluence case, therefore neglecting SSA. Further, the following parameters are used: $k_\mathrm{f,LE}=k_\mathrm{f,CT}=\frac{1}{1000}\,\mathrm{ps}^{-1}$, $D=10^{-2}\,\frac{\mathrm{cm}^2}{\mathrm{s}}$, $k_\mathrm{diss,CT} = 0.1\,\mathrm{ps}^{-1}$, $d=1\,$nm, $\lambda=0.45\,$eV, $|H_\mathrm{DA}| = 0.01\,$eV and $T=293.15\,$K. \\
	The presence of a substantial offset enables fast exciton dissociation, which in turn facilitates the formation of a significant concentration  gradient. This results in fast and efficient charge generation. Conversely, for offsets close to zero, the formation of a concentration gradient and the associated flux to the interface are inhibited. This is the reason why the characteristic GSB time in the main text is considerably longer than the average exciton diffusion time.}
	\label{fig:sup_concentration_profile}
\end{figure}

\newpage
\printbibliography

\makeatletter\@input{aux_alternative_main.tex}\makeatother